\def\pmb#1{\setbox0=\hbox{#1}%
\kern-.025em\copy0\kern-\wd0
\kern-.05em\copy0\kern-\wd0
\kern-.025em\raise.0433em\box0}

\def    \bOmega {\vec \Omega}
\def      \be       {\begin{equation}}
\def      \ee       {\end{equation}}

\documentclass{elsart}
{}

\usepackage{graphicx}

\usepackage{amssymb}
\newcommand{\Omegabold}{\mbox{{\boldmath $\vec \Omega$}}}
\begin{document}

\begin {frontmatter}
\title{Grain Alignment and CMB Polarization Studies}
\author[Alex]{A. Lazarian},
\address[Alex]{University of Wisconsin-Madison,
Astronomy Department, 475 N. Charter St., Madison, WI 53706, e-mail:
lazarian@astro.wisc.edu}

\begin{abstract}
Polarized microwave emission from dust is an important foreground that
may contaminate polarized CMB studies unless carefully accounted
for. Modeling of polarization from dust requires a quantitative understanding
of grain alignment. I review the current status of grain alignment theory outlining
recent advances in quantitative description of the alignment. In particular, I show
that the grain-alignment theory is a predictive one, and its results nicely match
 observations. Those indicate that the most important process of alignment is related to
 radiative torques acting on irregular grains. The recently developed analytical  model of
 radiative torque alignment has proven to be a very efficient tool for predicting the degree 
 of grain alignment. We expect the alignment theory to further mature before CMBPol flight, which
 would ensure a better accounting for the dust-related polarization. At the same time,
 CMBPol should provide the additional testing of grain alignment, clarifying the reliability of polarimetry
 for tracing of magnetic field.  
\end{abstract}
\end{frontmatter}

\section{Introduction}

Diffuse Galactic microwave emission, apart from carrying important information on
the fundamental
properties of the interstellar medium, interferes with
Cosmic Microwave Background (CMB) experiments
(see Bouchet et al. 1999). 
Polarization of the CMB provides information about the Universe
that is not contained in the temperature data alone. In particular, it
offers a unique
way to specifically trace the primordial perturbations of tensorial nature
({\em i.e.} cosmological gravitational waves),
and allows one to break some important degeneracies that remain in the 
measurement of cosmological
parameters with intensity alone (see Zaldarriaga et al. 1997).

Among different sources of polarized foregrounds, interstellar dust is probably
the most difficult to deal with (see Lazarian \& Finkbeiner 2003 for  a review). Polarized radiation from dust arises from grain being
aligned\footnote{In the absence of the alignment a marginal degree of polarization
can arise from difference of grain temperatures stemming from the difference in
grain cross sections in respect to the anisotropic radiation. In what follows, we disregard
the latter effect.}. Grain
alignment has proven to be a very difficult problem. It is well known that
the degree of grain alignment depends both on grain size and composition,
leading to the frequency dependence of the polarization (see Kim \& Martin 1994,
Hildebrand et al. 2000).  Moreover, in addition to the ``classical''
power-law distribution of larger grains (Mathis, Rumpl \& Nordsieck 1977), 
dust has a population of tiny grains (Leger \&
Puget 1984), which are frequently called PAH.  While the studies of
the alignment of such a particles, which are essentially large molecules, is only starting,
the theory of alignment of "classical" large grains has a long and exciting history (see
Lazarian 2003 for a review). 

Theory of grain alignment has been developing
fast recently, which makes even the most recent review by Lazarian (2007) outdated in a number of
respects. In what follows we provide an updated outlook on the grain alignment theory keeping in mind two
interrelated issues, namely,\\
a) how grain alignment theory can help in understanding of polarization to be measured by CMB-dedicated missions, e.g. CMBPol;\\
b) how CMBPol mission can help in understanding of grain alignment.

There are historic reasons why the researchers may be reluctant to rely on grain alignment theory
while modeling polarization from dust.  Grain alignment has a reputation of being an astrophysical problem of the longest standing.
Indeed, for a long time since the discovery of dust-induced starlight extinction polarization in 1949 (Hall 1949; Hiltner 1949) the mechanism of alignment was both enigmatic
and illusive. Thus, 
for many years grain alignment theory did not have predictive powers. Works by great minds like Lyman Spitzer
and Edward Purcell moved the field forward in terms of understanding the basic grain dynamics physics. However,
the theory was constantly failing to account for new observational data. Fig.~1a demonstrates the complexity of grain motion as we understand it now.

The weakness of the theory
 caused a somewhat cynical approach to it among some of the
polarimetry practitioners who preferred to be guided in their work
by the following rule of thumb: {\it All grains are always aligned
and the alignment happens with the longer grain axes perpendicular to
magnetic field.} This simple recipe was shattered, however, by observational
data which indicated that \\
I. Grains of sizes smaller than a critical size are either not aligned
or marginally aligned (Mathis 1986, Kim \& Martin 1995).\\
II. Carbonaceous grains are not aligned, while silicate grains are aligned
(see Mathis 1986).\\
III. A substantial part of small grains
grains deep within molecular clouds are not aligned (Goodman et al. 1995,
Lazarian, Goodman \& Myers 1997, Cho \& Lazarian 2005 and references therein).\\

These facts were eloquent enough to persuade even the most sceptical types
that the interpretation of interstellar polarimetric data does require an adequate
theory. The challenges related to separating of the contribution from dust from the 
one by polarized CMB makes it even more urgent to develop a predictive theory  of
grain alignment.

One should note that adequate grain alignment theory is able to give boost to many
astrophysical fields. First of all, polarization from aligned dust is a widely used way
to study magnetic fields in interstellar medium and molecular clouds. Needless to say that
ambiguities in grain alignment result in ambiguities of the interpretation of polarization
in terms of the underlying magnetic fields. Moreover, there are  numerous
astrophysical conditions when dust is bound to be aligned, but this fact is under-appreciated. For instance,
a common explanation of light polarization from comets or
circumstellar regions is based on light scattering by randomly oriented
particles (see Bastien 1988 for a review). The low efficiency and slow rates
of alignment were sometimes quoted to justify such an approach. However, it
has been proven recently that grain alignment is an efficient and rapid
process. Therefore, we {\it do expect} to have circumstellar, interplanetary,
and cometary dust aligned. Of particular interest in this respect are T-Tauri accretion
disks (see Cho \& Lazarian 2007). Tracing magnetic fields in these
environments with aligned grains opens new exciting avenues for polarimetry.

In this review I shall show that the modern grain alignment theory is consistent with the observations available.Moreover, the theory allows us for the first time ever make quantitative
{\it predictions} of the polarization degree from various astrophysical objects. However, the latter part of
the studies is in its infancy. We expect that the growing interest
to dust polarization from the CMB community will stimulate more work both in grain alignment theory and its observational testing. 
We feel that the CMBpol mission should both benefit from our current better understanding of grain alignment and provide
the critical testing of the remaining controvercies related to the grain alignment theory. 

Below, in \S 2 I shall show how the properties of polarized radiation are
related to the statistics of aligned grains. In \S 3 I shall briefly discuss the main
alignment mechanisms. In \S 4 I shall discuss the effects that are related to
internal dissipation of energy in a wobbling grains. In \S 5 I shall 
discuss the predictions obtained for the dominant mechanism of alignment, namely,
the radiative torque alignment. In \S 6, I shall discuss the prospects of grain alignment theory and 
 in \S 7 I shall discuss the synergy of studying of CMB polarization and improving of our knowledge of
 grain alignment.

\section{Aligned Grains \& Polarized Radiation}

A practical interest in aligned grains arises from the fact that their
alignment results in polarization of the starlight passing through dust as well as in
polarization in grain emission. Below we discuss why this happens.

\subsection{Polarized Starlight from Aligned Grains}

For an ensemble of aligned grains the degrees of extinction in the directions
perpendicular and parallel to the direction of alignment are
different\footnote{According to Hildebrand \& Dragovan (1995), the best fit of
the grain properties corresponds to oblate grains with the ratio of axis about
2/3.}. Therefore initially unpolarized starlight acquires polarization
while passing through a volume with aligned grains (see Fig.~2a). If the
extinction in the direction of alignment is $\tau_{\|}$ and in the
perpendicular direction is  $\tau_{\bot}$ one can write the polarization,
$P_{abs}$, by selective extinction
 of grains
as
\begin{equation}
P_{abs}=\frac{e^{-\tau_{\|}}-e^{-\tau_{\bot}}}{e^{-\tau_{\|}}+e^{-\tau_{\bot}}}
\approx -{(\tau_{\|}-\tau_{\bot})}/2~,
\label{Pabs}
\end{equation}
where the latter approximation is valid for $\tau_{\|}-\tau_{\bot}\ll 1$.
To relate the difference of extinctions to the properties of aligned grains
one can take into
account the fact that the extinction is proportional to the product
of the grain density and  their cross sections. If a cloud is composed of
identical aligned grains
$\tau_{\|}$ and $\tau_{\bot}$ are proportional to the number of grains
along the light path times the corresponding cross sections, which
are, respectively,
$C_{\|}$ and $C_{\bot}$.

In reality one has to consider additional complications (like, say, incomplete
grain alignment) and variations in the direction of the alignment axis
relative to the line of sight. (In most cases the alignment axis coincides
with the direction of magnetic field.) To obtain an adequate description, one
can (see Roberge \& Lazarian 1999) consider an electromagnetic wave
propagating along the line of sight (the {\mbox{$\hat{\bf z}^{\bf\rm o}$}}
axis, as on Fig.~1b). The transfer equations for the Stokes parameters depend
on the cross sections  $C_{xo}$ and $C_{yo}$ for linearly polarized waves with
the electric vector,  {\mbox{\boldmath$E$}}, along the {\mbox{$\hat{\bf
x}^{\bf\rm o}$}} and {\mbox{$\hat{\bf y}^{\bf\rm o}$}} axes perpendicular to
{\mbox{$\hat{\bf z}^{\bf\rm o}$}} (Lee \& Draine 1985).

\begin{figure}[h]
\includegraphics[width=4.in]{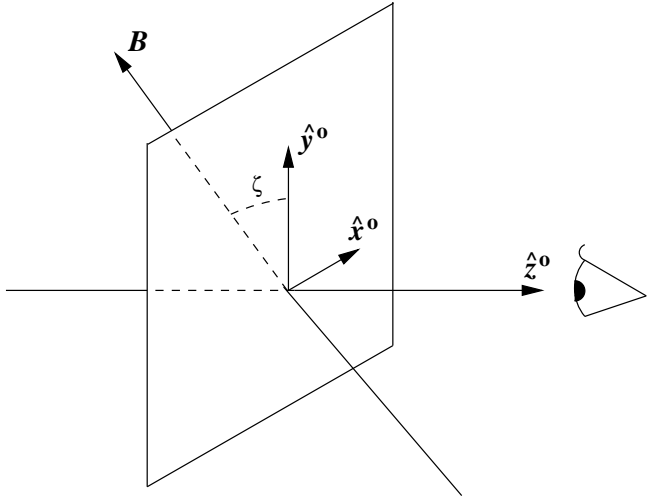}
\caption{\small 
 {\it (b) Right panel}--
Geometry of observations (after Roberge \& Lazarian 1999).}
\label{fig:2Dspek}
\end{figure}

To calculate  $C_{xo}$ and $C_{yo}$, one transforms the components of
{\mbox{\boldmath$E$}} to the principal axes of the grain, and takes the
appropriately-weighted sum of the cross sections $C_{\|}$ and  $C_{\bot}$ for
{\mbox{\boldmath$E$}} polarized along the grain axes (Fig~1b illustrates the
geometry of observations). When the transformation is carried out and the
resulting expressions are averaged over the precession angles, one finds (see
transformations in Lee \& Draine 1985 for spheroidal grains, and in Efroimsky
2002a for the general case) that the mean cross sections are
\begin{equation}
C_{xo} = C_{avg} + \frac{1}{3}\,R\,\left(C_{\bot}-C_{\|}\right)\,
       \left(1-3\cos^2\zeta\right)~~~,
\label{eq-2_5}
\end{equation}
\begin{equation}
C_{yo} = C_{avg} + \frac{1}{3}\,R\,\left(C_{\bot}-C_{\|}\right)~~~,
\label{eq-2_6}
\end{equation}
$\zeta$ being the angle between the polarization axis and the {\mbox{$\hat{\bf
x}^{\bf\rm o}$}} {\mbox{$\hat{\bf y}^{\bf\rm o}$}} plane, and
$C_{avg}\equiv\left(2 C_{\bot}+ C_{\|}\right)/3$ being the effective cross
section for randomly-oriented grains. To characterize the alignment, we used
in eq.~(\ref{eq-2_6}) the Rayleigh reduction factor (Greenberg 1968) defined
as
 \begin{equation}
 R\equiv \langle G(\cos^2\theta) G(\cos^2\beta)\rangle\;\;\;,
 \label{R}
 \end{equation}
 where angular brackets denote ensemble averaging, $G(x) \equiv 3/2 (x-1/3)$,
 $\theta$ is the angle between the axis of the largest moment of inertia
 (henceforth the axis of maximal inertia) and the magnetic field ${\bf B}$, while
 $\beta$ is the angle between the angular momentum ${\bf J}$ and ${\bf B}$.
 To characterize the alignment with respect to the magnetic field,
 the measures ${Q_X\equiv \langle G(\theta)\rangle}$ and $Q_J\equiv \langle G(\beta)
 \rangle$ are employed. Unfortunately, these statistics are not independent and
 therefore $R$ is not equal to $Q_J Q_X$ (see Lazarian 1998, Roberge \& Lazarian 1999).
 This considerably complicates the description of the alignment process.

\subsection{Polarized Emission from Aligned Grains}
Aligned grains emit polarized radiation (see Fig.~2b). The difference in
$\tau_{\|}$ and $\tau_{\bot}$ for aligned dust results in the emission
polarization:
\begin{equation}
P_{em}=\frac{(1-e^{-\tau_{\|}})-(1-e^{-\tau_{\bot}})}{(1-e^{-\tau_{\|}})+
(1-e^{-\tau_{\bot}})}\approx \frac{\tau_{\|}-\tau_{\bot}}
{\tau_{\|}+\tau_{\bot}}~,
\label{Pem}
\end{equation}
where both optical depths $\tau{\|}$ are $\tau_{\bot}$ were assumed to be
small. Taking into account that both $P_{em}$ and $P_{abs}$ are functions of
the wavelength $\lambda$ and combining eqs.(\ref{Pabs}) and (\ref{Pem}), one
obtains for $\tau=(\tau_{\|}+\tau_{\bot})/2$
\begin{equation}
P_{em}(\lambda) \approx -P_{abs}(\lambda)/\tau(\lambda)~,
\label{Pem}
\end{equation}
which establishes the relation between the polarizations in emission and
absorption. The minus sign in eq~(\ref{Pem}) reflects the fact that emission
and absorption polarizations are orthogonal. This relation enables one to predict the
far infrared polarization of emitted light if the starlight polarization is
measured. This opens interesting prospects of predicting the foreground polarization
arising from emitting dust using the starlight polarization measurements (Cho \& Lazarian
2002, 2003).
 As $P_{abs}$ depends on $R$,
$P_{em}$ also depends on the Rayleigh reduction factor.
\begin{figure}[h]
\includegraphics[width=3.in]{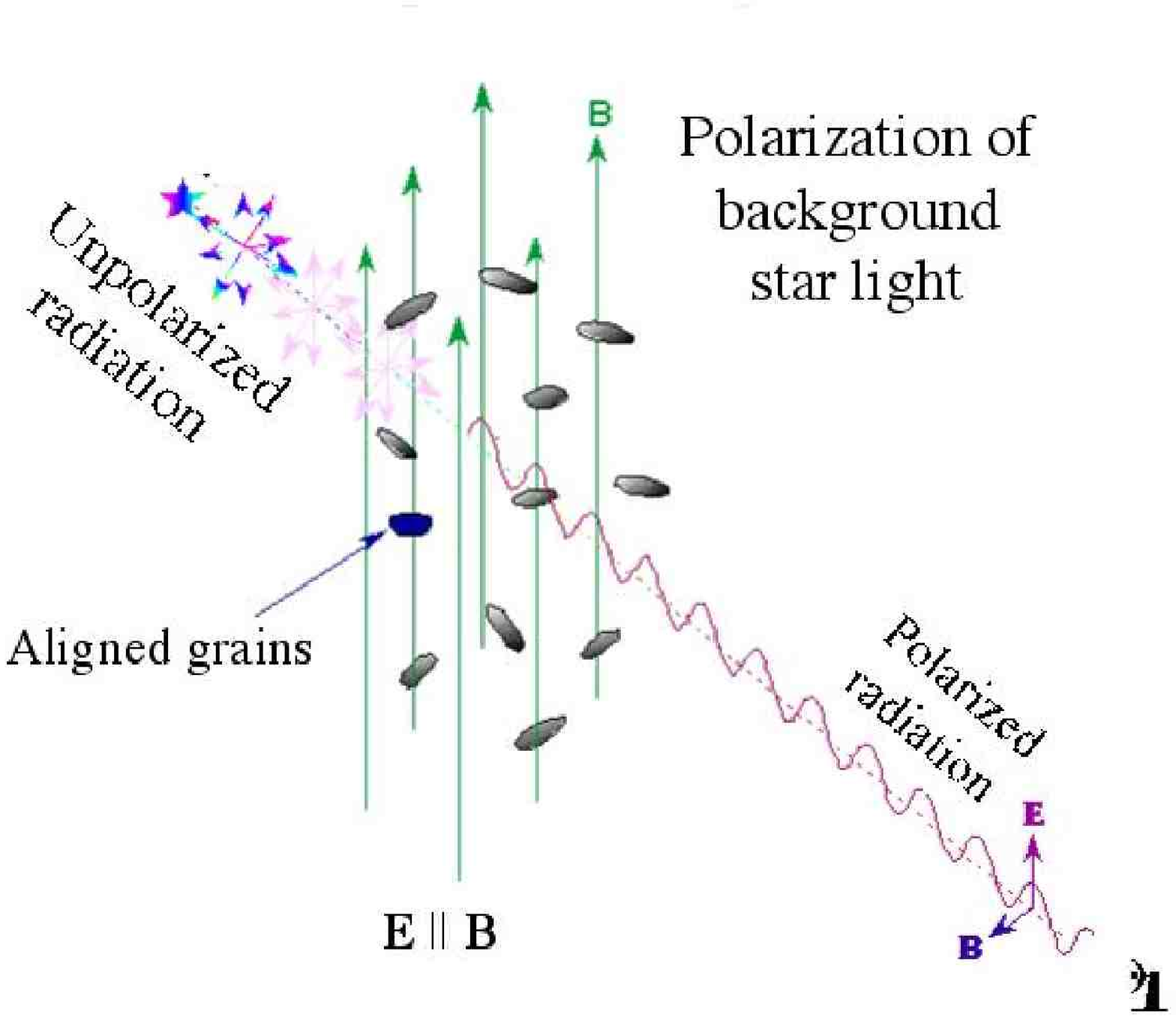}
\hfill
\includegraphics[width=3.in]{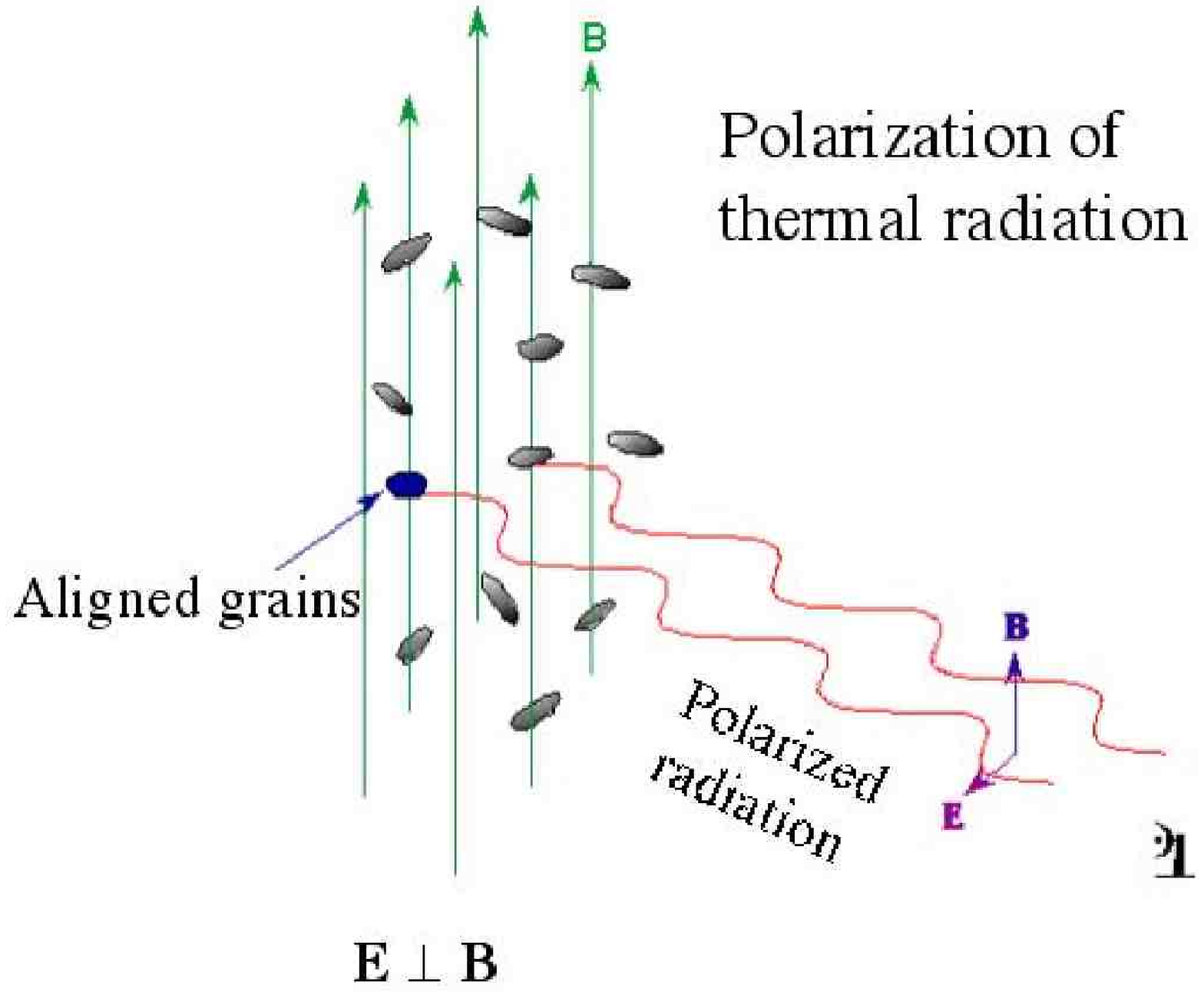}
\caption{
\small
{\it (a)Left panel}-- Polarization of starlight passing through a
cloud of aligned dust grains. The direction of polarization (${\bf E}$)
is parallel to the plane of the sky direction of magnetic field.
 {\it (b) Right panel}-- Polarization of radiation from a optically thin
cloud of aligned dust grains. The direction of polarization (${\bf E}$)
is perpendicular to the plane of the sky direction of magnetic field.
}
\label{fig:2Dspek}
\end{figure}

\subsection{Circular Polarization from Aligned Grains}
\label{circular}

One way of obtaining circular polarization is to have a magnetic field that
varies along the line of sight (Martin 1972). Passing through one cloud with
aligned dust the light becomes partially linearly polarized. On passing the
second cloud with dust gets aligned in a different direction. Hence, the light
gets circular polarization. Literature study shows that this effect is well
remembered (see Menard et al 1988), while another process entailing circular
polarization is frequently forgotten. We mean the process of single scattering
of light on aligned particles. An electromagnetic wave interacting with a
single grain coherently excites dipoles parallel and perpendicular to the
grain's long axis. In the presence of adsorption, these dipoles get phase
shift, thus giving rise to circular polarization. This polarization can be
observed from an ensemble of grains if these are aligned. The intensity of
circularly polarized component of radiation emerging via scattering of
radiation with ${\bf k}$ wavenumber on small ($a\ll \lambda$) spheroidal
particles is (Schmidt 1972)
\begin{equation}
V( {\bf e}, {\bf e}_0, {\bf e}_1)=\frac{I_0 k^4}{2 r^2}i(\alpha_{\|}
\alpha^{\ast}_{\bot}-\alpha^{\ast}_{\|}\alpha_{\bot})\left([{\bf e_0}\times
{\bf e}_1] {\bf e}\right)({\bf e}_0 {\bf e})\;\;\;,
\end{equation}
where ${\bf e}_0$ and ${\bf e}_1$ are the unit vectors in the directions of
incident and scattered radiation, ${\bf e}$ is the direction along the aligned
axes of spheroids; $\alpha_{\bot}$ and $\alpha_{\|}$ are the particle
polarizabilities along ${\bf e}$ and perpendicular to it.

The intensity of the circularly polarized radiation scattered in the volume
$\Delta \Gamma({\bf d}, {\bf r})$ at $|{\bf d}|$ from the star at a distance
$|{\bf r}|$ from the observer is (Dolginov \& Mytrophanov 1976)
\begin{equation}
\Delta V ({\bf d}, {\bf r})=\frac{L_{\star} n_{\rm dust}\sigma_{V}}{6\pi |{\bf d}|^4
|{\bf r}||{\bf d}-{\bf r}|^2}R \left([{\bf d}\times {\bf r}] h\right)
({\bf d}{\bf r})\Delta \Gamma({\bf d}, {\bf r})~~~,
\label{circular}
\end{equation}
where $L_{\star}$ is the stellar luminosity, $n_{\rm dust}$ is the number of
dust grains per a unit volume, and $\sigma_V$ is the cross section for
producing circular polarization, which for small grains is
$\sigma_V=i/(2k^4)(\alpha_{\|}\alpha^{\ast}_{\bot}-\alpha^{\ast}_{\|}\alpha_{\bot})$.
According to Dolginov \& Mytrophanov (1976) circular polarization arising from
single scattering on aligned grains can be as high as several percent for
metallic or graphite particles, which is much more than one may expect from
varying magnetic field direction along the line of sight (Martin 1971). In the
latter case, the linear polarization produced by one layer of aligned grains
passes through another layer where alignment direction is different. If on
passing through a single layer, the linear polarization degree is $p$, then
passing through two layers produces circular polarization that does not exceed
$p^2$.

\section{Grain Alignment Theory: Major Mechanisms}

\subsection{Characteristic time scales involved in grain alignment}

We have seen in the previous sections that both linear and circular
polarizations depend on the degree of grain alignment given by the
Rayleigh reduction
factor (see Eq.~(\ref{R})). Therefore it is the goal of the grain alignment theory
to determine this factor. Table~1 shows that the wide range of different time
 scales involved makes the brute force numerical approach doomed.

\begin{figure}
\includegraphics[width=2.2in]{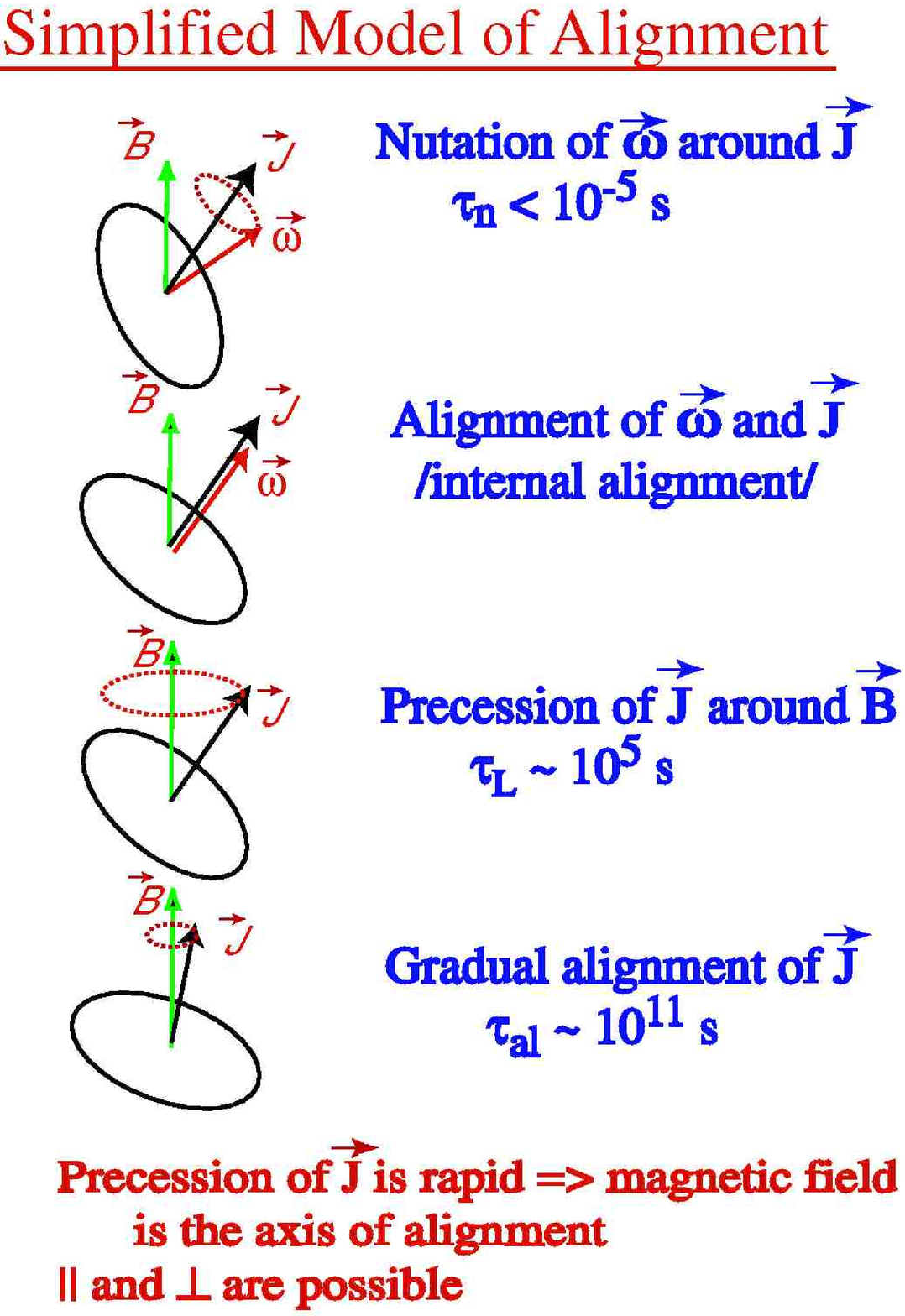}
\hfil
\includegraphics[width=3.8in]{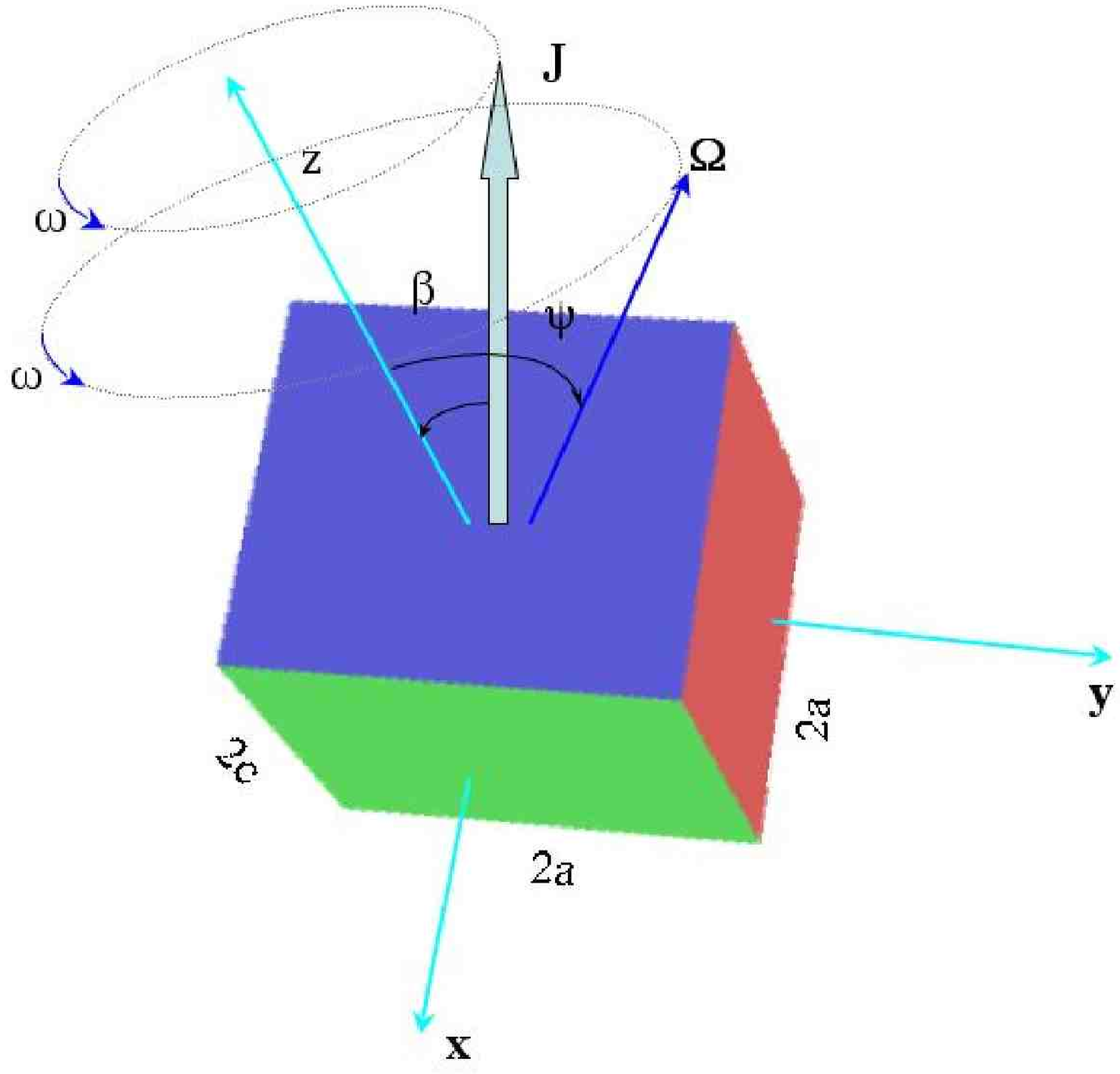}
\caption{\small{\it Left panel}-- Alignment of grains implies several
alignment processes acting simultaneously and covering various timescales.
Internal alignment was introduced by Purcell (1979) and was assumed to be a
slow process. Lazarian \& Draine (1999a) showed that the internal alignment is
$10^6$ times faster if nuclear spins are accounted for.  The time scale of
${\bf J}$ and ${\bf B}$ alignment is given for diffuse interstellar medium. It
is faster in circumstellar regions and for cometary dust. {\it Right panel}-- Grain dynamics as seen in the grain
frame of reference. The Barnett magnetization is directed along $\bOmega$, and
it causes a gradual grain remagnetization as the $\bOmega$ precesses in the
grain axes. }
\label{grain}
\end{figure}

Grain motion as it undergoes alignment is rather complex (see Fig.~\ref{grain}). This makes
the problem of grain alignment very difficult.

A number of different mechanisms that produce grain alignment has been
developed by now. Dealing
with a particular situation one has to identify the dominant alignment process.
Therefore it is essential to understand different mechanisms.

\begin{table}[h]
\caption{Time-scales relevant for grain alignment}
\begin{displaymath}
\begin{array}{rrrrrr} \hline\hline\\
\multicolumn{1}{c}{\rm Symbol} & \multicolumn{1}{c}{\rm Meaning}  & \multicolumn{1}{c}{\rm Definition} &
\multicolumn{1}{c}{\rm Value~~(s)} \\[1mm]
\hline\\
{\rm t_{rot}}&{\rm thermal~rotational~period}&{2\pi/\Omega}
&{6\times 10^{-5}\hat{T}_{rot}^{-1/2}a_{-5}^{5/2}s^{-2}}\\[1mm]
{\rm t_{Bar}}&{\rm Barnett~relaxation~time}&
{\frac{\gamma_e^{2}I_{\|}^{3}}{VK_{e}h^{2}(h-1)J^{2}}}&{9.84\times 10^{6}(\frac{\hat{\rho}^{2}}{\hat{K_{e}}\hat{T}_{d}})f_1(s) a_{-5}^{7}}(\frac{J_{d}}{J})^{2}F(\tau_{el})\\[1mm]
{\rm t_{nucl}}&{\rm
  nuclear~relaxation~time}&{(\frac{\gamma_n}{\gamma_e})^2(\frac{K_e}{K_n}) t_{Bar}}
  &{21.35 \hat{\rho}^{2} a_{-5}^{7}f_{1}(s)\hat{g}_{n}^{4}\hat{\mu}_n^{-2}}(\frac{J_{d}}{J})^{2}F(\tau_n)\\[1mm]
{\rm t_{c}}&{\rm crossover~time}&{\frac{2 J_{d,\bot}}{L_{z}^{b}}~~~~} & {1.6\times 10^{9} (\frac{\hat{\rho}
\hat{T_d}\hat{\alpha}}{\hat{W}\hat{\zeta}^2\hat{n}^2 \hat{T}_{g}})^{1/2} f_2(s) a_{-5}^{1/2}} \\[1mm]
{\rm t_{L}} &{\rm Larmor~precession~time}& {\frac{2\pi\mu_{d} I_{\|}}{\chi^{'}VB}}&{1.1\times
   10^{6}(\frac{\hat{\rho}\hat{T_{d}}}{\hat{\chi}\hat{B}})a_{-5}^{2}}s^{2}\\[1mm]
{\rm t_{RT}}&{\rm Radiative~precession~time}&{\frac{2\pi}{|d\phi/dt|}}&{\frac{3\times 10^{7}}{\hat{Q}_{e3}}\hat{\rho} b_{-5}^{1/2}(\frac{1}{\hat{\lambda}\hat{u}_{rad}})} \\[1mm]
{\rm t_{gas}}&{\rm gas~damping~time}& {\frac{4I_{\|}}{nmv_{th} b^{4}}}& {4.6\times
  10^{12}(\frac{\hat{\rho}_{s}}{\hat{n}\hat{T}_{g}^{1/2}}) sb_{-5}}\\[1mm]
{\rm t_{E}}&{\rm electric~precession~time}& {\frac{2\pi}{\Omega_{E}}}&{0.2 \times 10^{11} p^{-1} \hat{E}^{-1}\hat\rho \hat{\omega}{a_{-5}}}\\[1mm]
{\rm t_{DG}}&{\rm paramagnetic~damping~time}&{\frac{2\rho a^{2}}{5K(\omega)T_{2}B^{2}}}&{10^{13}\hat{B}^{-1}\hat{K}^{-1}a_{-5}^{2}s^2}\\[1mm]
\hline
\end{array}
\end{displaymath}
{\tiny Notations:\\
$a$: minor axis~ ~~~~~~~~~~~~~~~~~~~~~~~~~~~~~~~~~~~~~~~~~~~~~~~~~~~~~~~~~~~~~~~$b$: major axis\\
$a_{-5}=a/10^{-5} cm$~~~~~~~~~~~~~~~~~~~~~~~~~~~~~~~~~~~~~~~~~~~~~~~~~~~~~~~~~~~~$s=a/b < 1$: ratio of axes\\
$h=I_{\|}/I_{\perp}$: ratio of moment inertia~~~~~~~~~~~~~~~~~~~~~~~~~~~~~~~~~~~~$\hat{\rho}=\rho/3 gcm^{-3}$: normalized grain density\\
$\hat{T}_{g}=T_{g}/85 K$: normalized gas
temperature~~~~~~~~~~~~~~~~~~~~~~~~~~$\hat{T}_{d}=T_{d}/15 K$: normalized dust
temperature\\
$T_{rot}=(T_{g}+T_{d})/2$: rotation temperature\\
$\hat{n}=n/20 cm^{-3}$: normalized gas
density~~~~~~~~~~~~~~~~~~~~~~~~~~~~~~$\hat{B}=B/5 \mu
\mbox{G}$: normalized magnetic field\\
$\chi^{'}=10^{-3}\hat{\chi}/\hat{T_{d}}$: real part of magnetic
susceptibility~~~~~~~~~~~~$\hat{K}_{e}=K_{e}/10^{-13}  F^{-1}(\tau_e)$\\
$K_{e,n}\omega$: imaginary part of magnetic susceptibility by electron and nuclear spin\\
$\mu_d$: grain magnetic
moment~~~~~~~~~~~~~~~~~~~~~~~~~~~~~~~~~~~~~~~~~~~~~~$\gamma_e=\frac{g_{e}\mu_{B}}{\hbar}$: magnetogyric ratio for electron\\
$\gamma_{n}=\frac{g_{n}\mu_{n}}{\hbar}$: magnetogyric ratio nuclei\\
$\hat{\mu}_n=\mu_n/\mu_N$: normalized magnetic moment of nucleus~~~~~~~~~~$\mu_N=e\hbar/2m_pc=5.05\times 10^{-24}$~ergs G$^{-1}$\\
$J_{d}=(\frac{I_{\|}I_{\perp}kT_{d}}{I_{\|}-I_{\perp}})^{1/2}$: grain angular 
momentum at  $T=T_{d}$ ~~~~~~~$J_{therm}$: grain angular momentum at $T=T_{gas}$\\
$t_{B,nucl}^{-1}=t_{B}^{-1}+t_{nucl}^{-1}$: total nuclear relaxation time~~~~~~~~~~~~ can also include
inelastic relaxation\\
$\hat{u}_{rad}=u_{rad}/u_{ISRF}$: energy density of radiation
field~~~~~~~~~~~$\hat{\lambda}=\overline{\lambda}/1.2\mu m$: wavelength of radiation field\\
$\hat{Q}_{e3}=\mbox{Q}_{\Gamma}.\mbox{e}_{3}/10^{-2}$: third component of radiative torques~~~~~~~$E=\hat{E}/10^{-5} Vcm^{-1}$: electric field\\
$p=10^{-15} \hat{U}a_{-5}\hat{\kappa_{e}}$: electric dipole moment~~~~~~~~~~~~~~~~~~~~~~~~~~$\hat{\kappa_{e}}=\kappa_{e}/10^{-2}$: electric constant\\
$\hat{U}=U/0.3 V$: normalized voltage~~~~~~~~~~~~~~~~~~~~~~~~~~~~~~~~~~~~~~~$\hat{\omega}=\omega/10^{5}
rad~s^{-1}$: angular velocity\\
$L_{z}^{b}$: magnitude of $H_{2}$
torque~~~~~~~~~~~~~~~~~~~~~~~~~~~~~~~~~~~~~~~~~~~~~$\hat{\zeta}=\zeta/0.2$
fraction of absorbed atoms\\
$\hat{W}=W/0.2$: kinetic energy of H$_2$~~~~~~~~~~~~~~~~~~~~~~~~~~~~~~~~~~~~~$\hat{\alpha}=\alpha/10^{11}$~cm$^{-2}$: density 
of recombination sites\\
$F(\tau)\equiv
[1+(\Omega\tau/2)^2]^2$~~~~~~~~~~~~~~~~~~~~~~~~~~~~~~~~~~~~~~~~~~~~~~~~~~$\tau_n$: nuclear
spin-spin relaxation rate\\
$\tau_{el}$: electron spin-spin relaxation rate~~~~~~~~~~~~~~~~~~~~~~~~~~~~~~~~$\mu_e\approx\mu_B$;
$\mu_B\equiv e\hbar/2m_ec$: Bohr magneton\\
$f_1(s)\equiv{s^{-6}(1+s^{2})^{2}}$~~~~~~~~~~~~~~~~~~~~~~~~~~~~~~~~~~~~~~~~~~~~~~~~~~~~~$f_2(s)\equiv(\frac{1+s^{2}}{s(1-s^{2})})^{1/2}$\\

}

\end{table}

\subsection{Paramagnetic Alignment}
\label{paramagnetic}

The Davis-Greenstein (1951)
mechanism (henceforth D-G mechanism)
is based on the paramagnetic dissipation that is experienced
by a rotating grain. Paramagnetic materials contain unpaired
electrons that get oriented by the interstellar magnetic field ${\bf B}$.
The orientation of spins causes
grain magnetization and the latter
varies as the vector of magnetization rotates
 in the grain body coordinates. This causes paramagnetic loses
at the expense of the grain rotation energy.
Be mindful, that if the grain rotational velocity ${\Omegabold}$
is parallel to ${\bf B}$, the grain magnetization does not change with time
and therefore
no dissipation takes place. Thus the
paramagnetic dissipation  acts to decrease the component of ${\Omegabold}$
perpendicular to ${\bf B}$ and one may expect that eventually
grains will tend to rotate with ${\Omegabold}\| {\bf B}$
provided that the time of relaxation $t_{D-G}$ is much shorter than
the
time of randomization through chaotic gaseous bombardment, $t_{gas}$.
In practice, the last condition is difficult to satisfy. It is clear from 
Table~1 that for $10^{-5}$ cm
grains
in the diffuse interstellar medium,
$t_{D-G}$ is of the order of $10^{13}a_{(-5)}^2 s^2 B^{-2}_{(5)}$s ,
while  $t_{gas}$ is $5\times 10^{12}n_{(20)}T^{-1/2}_{(2)} a_{(-5)}$ s if
magnetic field is $10^{-5}$ G and
temperature and density of gas are $100$ K and $20$ cm$^{-3}$, respectively.

The first detailed analytical treatment of the problem of D-G
alignment was given by Jones \& Spitzer (1967) who described the alignment
of ${\bf J}$
using the Fokker-Planck equation. This
approach allowed them to account for magnetization fluctuations
within the grain material, and thus provided a more accurate picture of the
${\bf J}$ alignment.
The first numerical treatment of
D-G alignment was presented by Purcell (1969).
By that time, it became clear that the original D-G
mechanism is too weak to explain the observed grain alignment. However,
Jones \& Spitzer (1967) noticed that if interstellar grains
contain superparamagnetic, ferro- or ferrimagnetic
inclusions\footnote{The evidence for such inclusions was found much later
through the study of interstellar dust particles captured in
the atmosphere (Bradley 1994).}, the
$t_{D-G}$ may be reduced by orders of magnitude. Since $10\%$ of
atoms in interstellar dust are iron,
the formation of magnetic clusters in grains was not far fetched
(see Martin 1995).
However, detailed calculations in Roberge \& Lazarian
(1999) showed that the degree of alignment achievable cannot account for the
observed polarization coming from molecular clouds if grains rotate thermally.
 This is the consequence of the
thermal suppression of paramagnetic alignment first discussed
by Jones \& Spitzer (1967). These internal
magnetic fluctuations
randomize grain orientation with respect to the magnetic field if the
grain body temperature is close to the rotational one.

Purcell (1979) pointed out that fast rotating grains are immune to
both gaseous and internal magnetic randomization. Substantial degrees of
alignment could be possible if the rotation in the same direction could
be preserved by the external uncompensated torques. Both the radiative
torques and other, subtle processes that we discuss dealing with grain 
rotation limit the ability of grain alignment in accordance with the Purcell (1979) suggestion (see
Spitzer \& McGlynn 1979, Lazarian \& Draine 1999a, Hoang \& Lazarian 2008).

Lazarian \& Draine (2000) predicted
that PAH-type particles can be aligned paramagnetically due to the relaxation
that is faster than the DG predictions. In fact, they showed that the DG
alignment is not applicable to very swiftly rotating particles, for which
the Barnett magnetic field gets comparable to magnetic fields induced by uncompensated
spins in the paramagnetic material. For such grains, this relaxation is more efficient than the
one considered  by Davis \& Greenstein (1951). This effect, that is termed ``resonance relaxation''
in Lazarian \& Draine  (2000), allows the alignment of PAHs. These tiny ``spinning'' grains
are responsible for the anomalous foreground
microwave emission (Draine \& Lazarian 1998,
see also Lazarian \& Finkbeiner 2003 for a review).

\subsection{Mechanical Alignment of Regular Grains: supersonic alignment}
\label{supersonic}

The Gold (1951) mechanism is a process of mechanical alignment of regular
grains. We shall understand by regular grains the grains that do not exhibit
helicity (cf. \S\ref{subsonic}). Grains presented by spheroids and bricks are the kind 
of regular grains we are talking about.

The original Gold (1951) idea is easy to understand. 
Consider
a needle-like grain interacting with a stream of atoms. Assuming
that collisions are inelastic, it is easy to see that every
bombarding atom deposits with the grain an angular momentum $\delta {\bf J}=
m_{atom} {\bf r}\times {\bf v}_{atom}$,
which is directed perpendicular to both the
needle axis ${\bf r}$ and the
 velocity of atoms ${\bf v}_{atom}$. It is obvious
that the resulting
grain angular momenta will be in the plane perpendicular to the direction of
the stream. It is also easy to see that this type of alignment will
be efficient only if the flow is supersonic\footnote{Otherwise grains
 see atoms coming not from one direction, but from a wide cone of
directions (see Lazarian 1997a) and the efficiency of
alignment decreases.}.

The quantitative
numerical study of the Gold alignment in Roberge et al. (1995) was
done under the assumption of  the perfect
coupling of ${\bf J}$ with the axis of maximal inertia.
This study shows a good
correspondence with an analytical formulae for the alignment
of ${\bf J}$ vector in L94 when
the gas-grain velocities are hypersonic. An analytical study in
Lazarian (1997) accounts for the incomplete internal alignment in
a more sophisticated way, compared to L94, and
predicts the Rayleigh reduction factors of $20\%$ and more for grains
interacting with the Alfven waves. 

As we discuss below, Purcell (1979) introduced the concept of pinwheel torques
that act to induce grain rotation with velocities much in excess of the
thermal Brownian rate. Such fast rotating grains seemed not to
be susceptible to the Gold (1951) process. This persuaded many researchers
that mechanical alignment is of marginal importance. Indeed, it seems natural to accept
that fast rotation makes
it difficult for gaseous bombardment to align grains. However, the
actual situation is more interesting. First of all, it was proven that
 mechanical alignment of suprathermally rotating grains
is possible (Lazarian 1995). Two mechanisms that were termed
``crossover'' and ``cross section'' alignment were introduced
there. The mechanisms were further elaborated and quantified
in Lazarian \& Efroimsky (1996), Lazarian, Ozik \& Efroimsky (1996),
Efroimsky (2002b). Second, Lazarian \& Drane (1999ab) showed that
grains may be thermally trapped and not rotate at high speed in spite of the
presence of pinwheel torques (see more in \S \ref{pinwheel}). The original Gold (1951) alignment
should be applicable to such grains. Moreover, supersonic velocities
may be available in the interstellar medium due to MHD turbulence (Lazarian \& Yan 2002,
Yan \& Lazarian 2003, Yan, Lazarian \& Draine 2004).

Therefore, the problem of the mechanical alignment of regular grains is not that 
it is not efficient, but that in many cases the process may be subdominant. This 
is the case, for instance, when a grain has intrinsic helicity. The mechanical alignment
may be important for small grains, e.g. PAHs, if they move supersonically in respect to
gas or flow of ions.

\subsection{Mechanical Alignment of Helical Grains: subsonic alignment}
\label{subsonic}

Astrophysical grains are known to be irregular. Therefore they can, for 
instance, scatter more impinging atoms to the right than to the left, the same
way that a model grain in Fig.~\ref{AMO} scatters photons. If present, the helicity radically
changes the dynamics of grains.

As far as we know, the mechanical alignment of helical grains was first briefly
discussed in Lazarian (1995) and
 Lazarian, Goodman \& Myers (1997), but was not elaborated there. Lazarian \& Hoang (2007b) 
 calculated the torques acting on a model  helical grains and showed that effective
 alignment is possible for subsonic drifts of grains in respect to the  ambient gas. The
 estimates obtained their are very encouraging, but they appeal to the details of
 grain-gas interaction that should be clarified by the further research.
 
 It is easy to understand why supersonic drift may not be necessary for helical
grains. For such grains 
the momentum deposited by regular torques scales in proportion to the number of collisions,
while the randomization adds up only as a random walk. In fact, the difference
between the mechanical alignment of spheroidal and helical grains is similar
to the difference between the Harwit (1971) alignment by stochastic absorption
of photons and the radiative torque alignment (see \S~\ref{RATs}). Indeed, while the
Harwit (1971) alignment requires very special
conditions to overpower the gaseous and emission randomization (Purcell \& Spitzer 1971), the radiative torques acting on a helical
grain easily beat randomization.

\subsection{Radiative Torque Alignment}
\label{RATs}

\begin{figure}[h]
\includegraphics[width=3.3in]{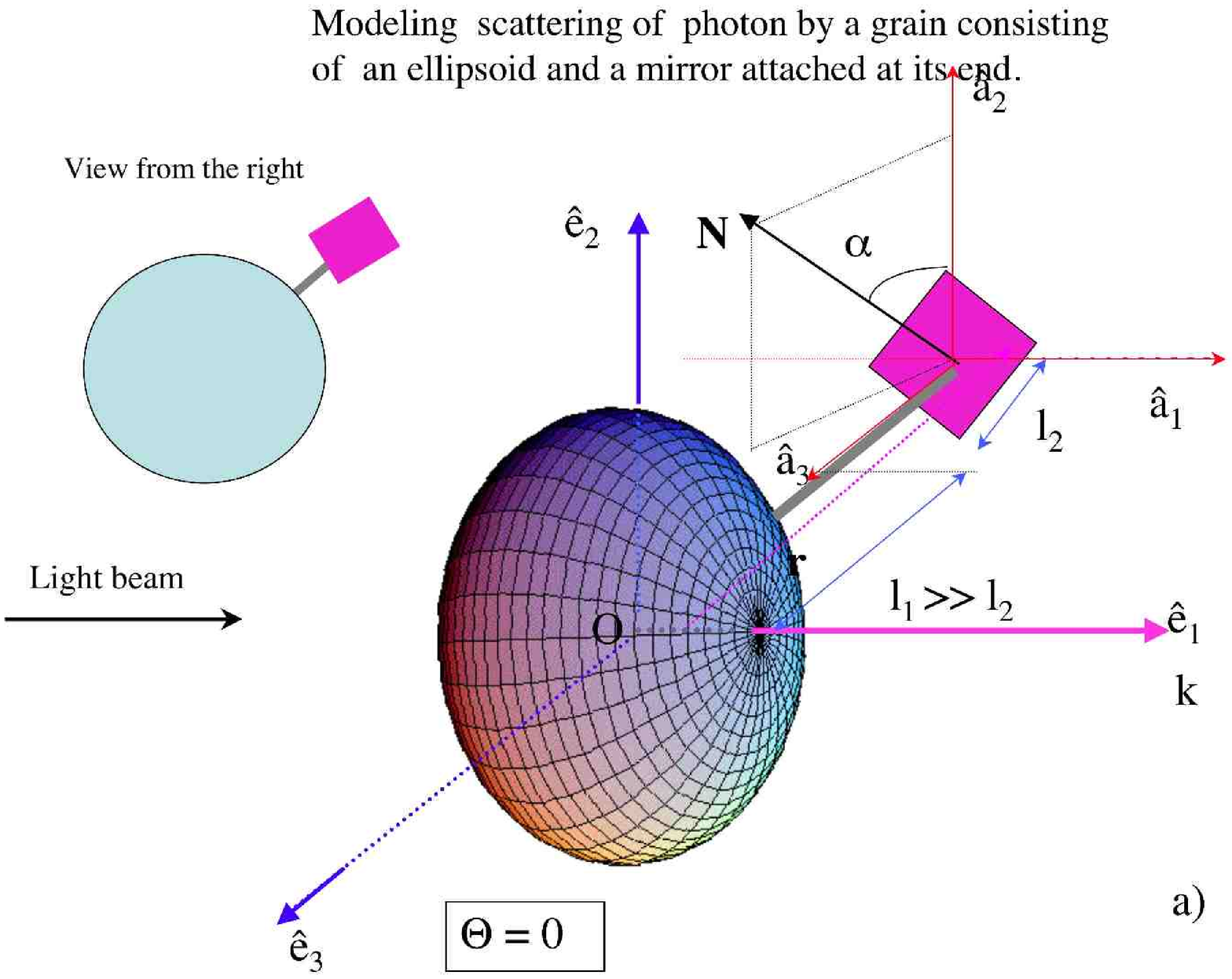}
\hfill
\includegraphics[width=2.5in]{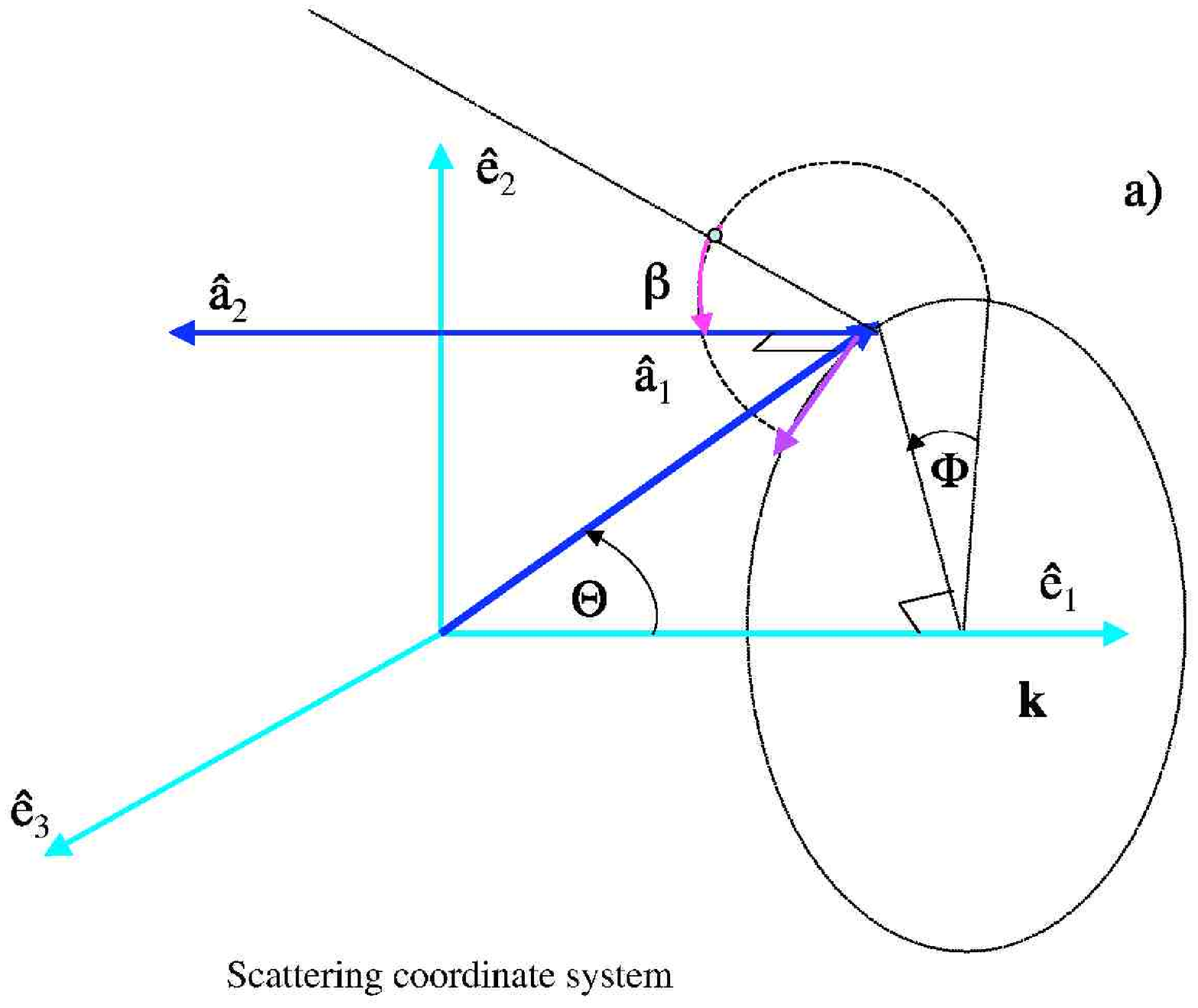}
\caption{
\small
{\it (a) Left panel}.-- A model of a ``helical'' grain,
that consists of a spheroidal grain with an inclined mirror attached to it,
reproduces well the radiative torques (from Lazarian \& Hoang 2007a).
 {\it (b) Right panel}.-- The ``scattering coordinate system'' which
illustrates the definition of torque components: ${\bf a_1}$ is directed
along the maximal inertia axis of the grain; ${\bf k}$ is the direction of radiation.
The projections of normalized radiative torques $Q_{e1}$,
$Q_{e2}$ and $Q_{e3}$ are calculated in this reference frame for $\Phi=0$.}
\label{AMO}
\end{figure}

{\it Stochastic} radiative torques arising from the absorption from a star were considered by Harwit (1971), who addressed the issue of
interaction of symmetric, e.g. spheroidal, grains with a radiative flow. These torques were shown to be subdominant in most
astrophysical conditions (Purcell \& Spitzer 1971). In what follows, we do not consider stochastic radiative torques. 

{\it Regular} radiative torques were introduced by Dolginov and Mytrophanov (1976) who considered ``twisted grains'' that can be
characterized by some {\it helicity}. They noticed that ``helical'' grains
would scatter differently the left- and right-polarized light, for which
reason an ordinary unpolarized light would spin them up.
        The subset of the ``helical'' grains was not properly identified,
and the later researchers could assume that it is limited to special
shapes/materials. One way or another, this ground-breaking work did not make
much impact to the field until Draine \& Weingartner (1996)
numerically showed that grains of rather arbitrary irregular shapes experience strong
radiative torques. This initially was interpreted as the evidence of the fast grain rotation, but
later research that radiative torques can both spin-up and slow down grains. In particular,
 Lazarian \& Hoang (2007a) noticed that grain rotation may
be essentially stopped by radiative torques. In the presence of thermal fluctuations grains do not completely 
stop, but rotate at subthermal rates at low-$J$ attractor points (Weingartner \& Draine 2003, Hoang \& Lazarian 2008a). 

A key point is that anisotropic starlight radiation can {\it align} grains on their own, i.e. without
the influence of any other, e.g. paramagnetic, mechanism. In fact, in most cases, in particular in 
diffuse interstellar gas, the radiative torques are so strong that they absolutely dominate the
dynamics of grains. 

The effect of alignment induced by radiative torques was discovered by  
Dolginov \& Mytrophanov (1976).  In their paper they
considered a tilted oblate grain with the helicity axes coinciding with
the axis of maximal inertia, as well as a tilted prolate grain for which
the two axes were perpendicular. They concluded, that subjected to a radiation
flux, the tilted oblate grain will be aligned with longer axes perpendicular
to magnetic field, while the tilted prolate grain will be aligned with the
longer axes parallel to magnetic field. 
The problem was revisited by Lazarian (1995), who took into account
the internal relaxation of energy within grains and concluded that both prolate  and oblate grains
will be aligned with longer axes perpendicular to the magnetic field. However,
Lazarian (1995) did not produce quantitative calculations and underestimated
the relative importance of radiative torque alignment compared to other
mechanisms. Curiously enough is that our recent analysis of the alignment of an irregular grain
mechanical properties of which are presented by a triaxial ellipsoid revealed that, 
even without internal relaxation, grains tend to get aligned with long axes perpendicular
to magnetic field (Hoang \& Lazarian 2008c).

The explosion of interest to the radiative torques we owe to
Bruce Draine, who realized that the torques
can be treated with the DDA
code by Draine \& Flatau (1994) and modified the code correspondingly.
The magnitude of torques were found to be substantial and present
for grains of all irregular shapes studied in Draine (1996), Draine \& Weingartner (1996, 1997). After that it became impossible to ignore the radiative torque alignment. More recently, radiative
torques have been studied in laboratory conditions (Abbas et al. 2004).
In what follows we identify the alignment by radiative torques with the alignment of grains by the
anisotropic radiation flux. This corresponds to the original thinking in Dolginov \& Mytorphanov (1976).

The Draine \& Weingartner (1996, 1997) papers signified a qualitative change in
the landscape of grain alignment theory. Although the numerical modeling  there was rather simplified\footnote{In fact, the crossovers induced by radiative troques were treated incorrectly in Draine \& Weingartner (1997), which induced a notion of {\it cyclic} trajectories. A rigorous treatment of crossovers
in Lazarian \& Hoang (2007a) corrected this point and showed that cyclic trajectories for grains do not exist. Instead, in the absence of gaseous bombardment, grains get settled either at an attractor point.}, these papers conjectured that radiative torques alignment may be the dominant mechanism in the diffuse
interstellar medium. Weingartner \& Draine (2003) attempted  to make more realistic modeling by accounting for thermal fluctuations within grains (see \S\ref{wobbling}).

\begin{figure}[h]
\includegraphics[width=1.7in]{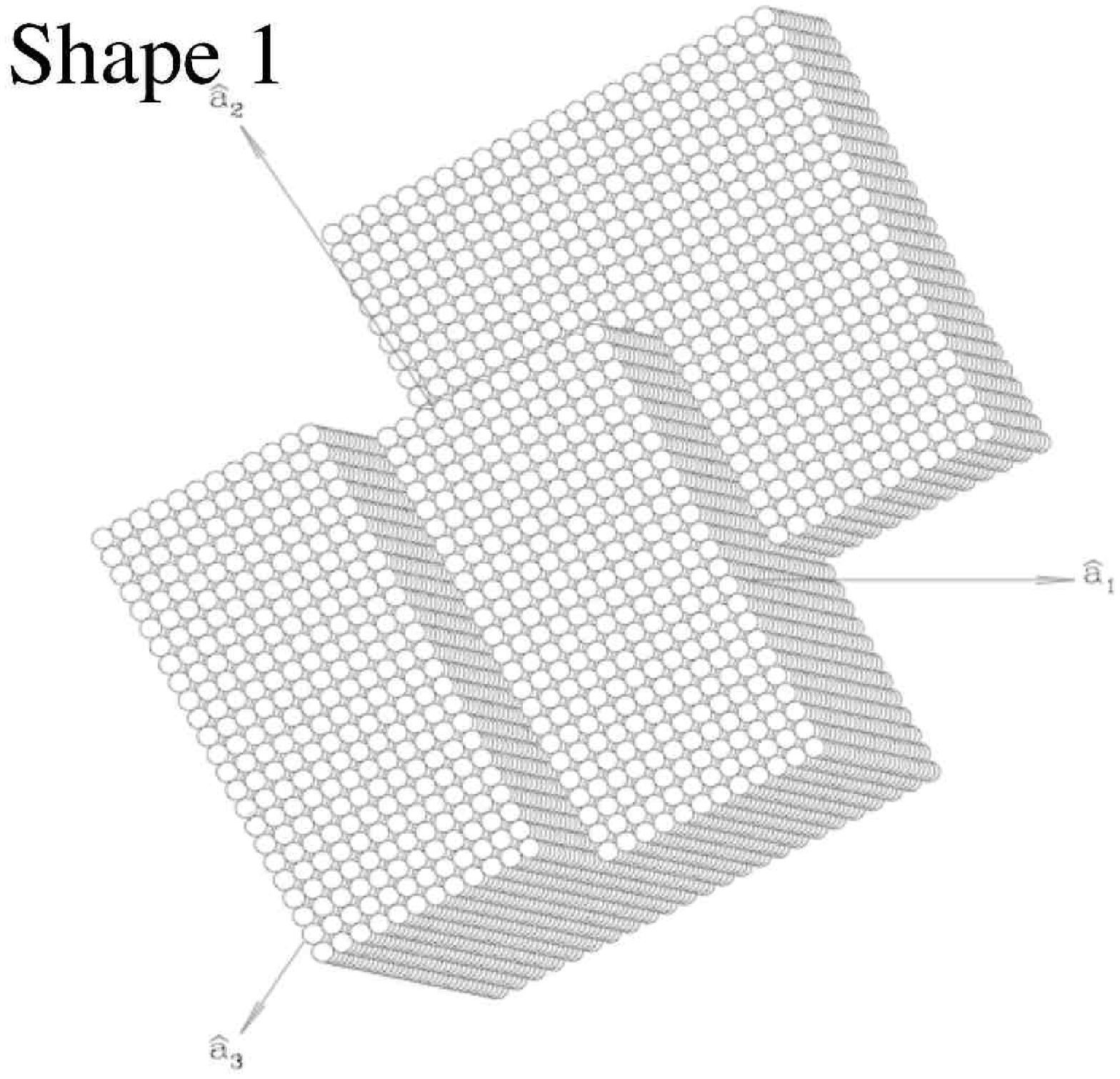}
\includegraphics[width=1.7in]{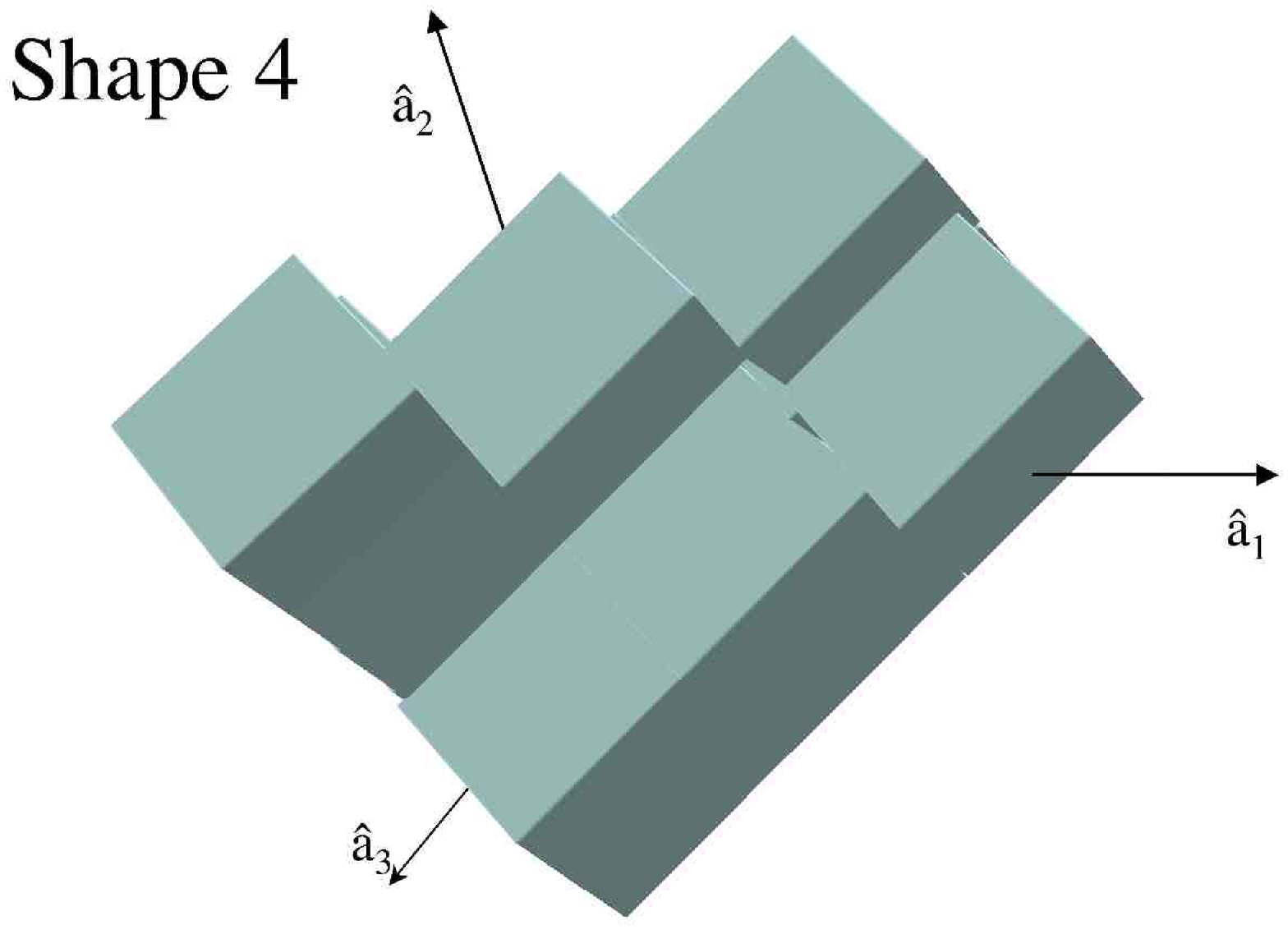}
\includegraphics[width=1.7in]{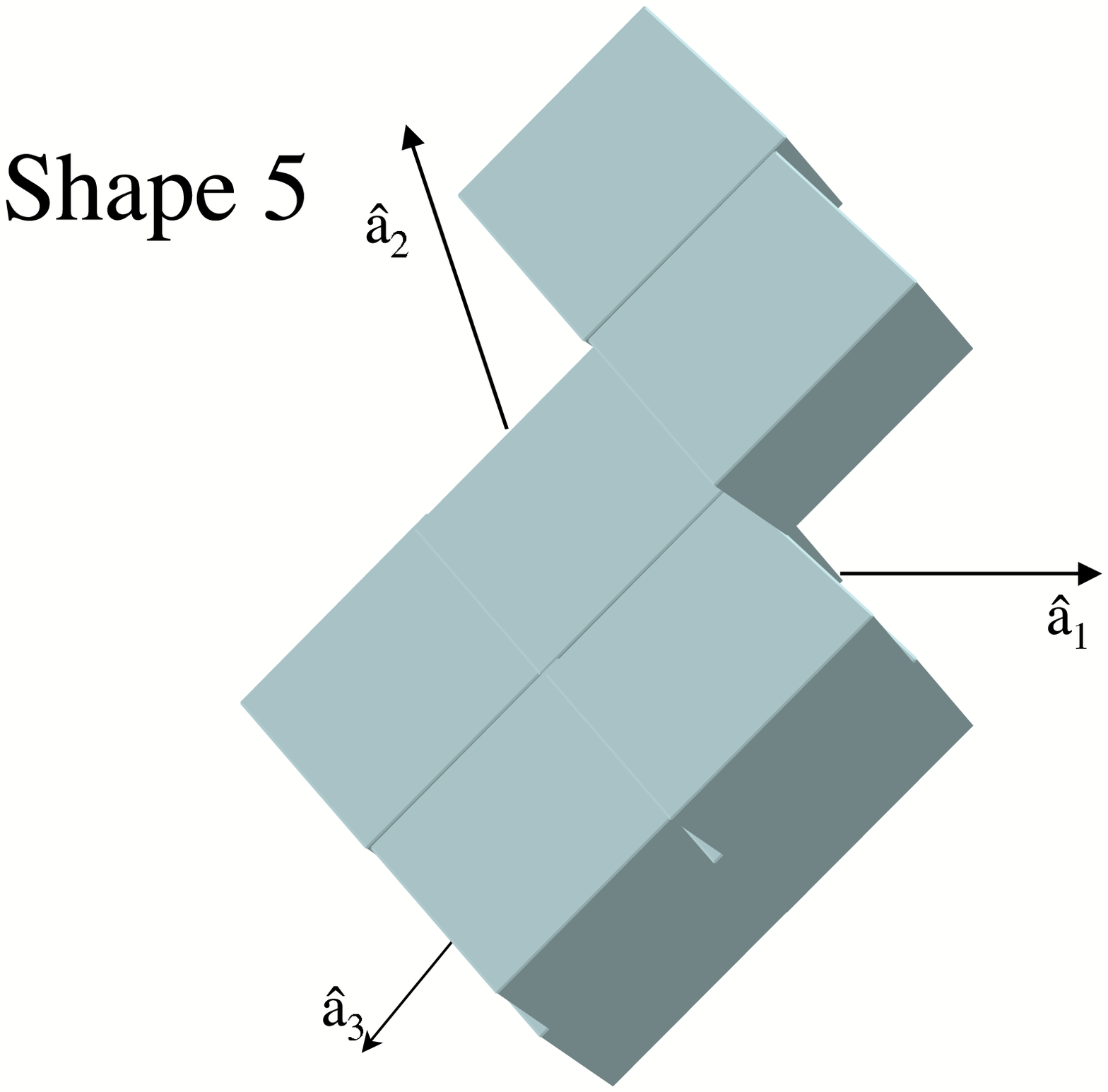}
\caption{\small Examples of irregular shapes studied in Lazarian \& Hoang 2007a.}
\label{shapes}
\end{figure}
\begin{figure}[h]
\includegraphics[width=3.in]{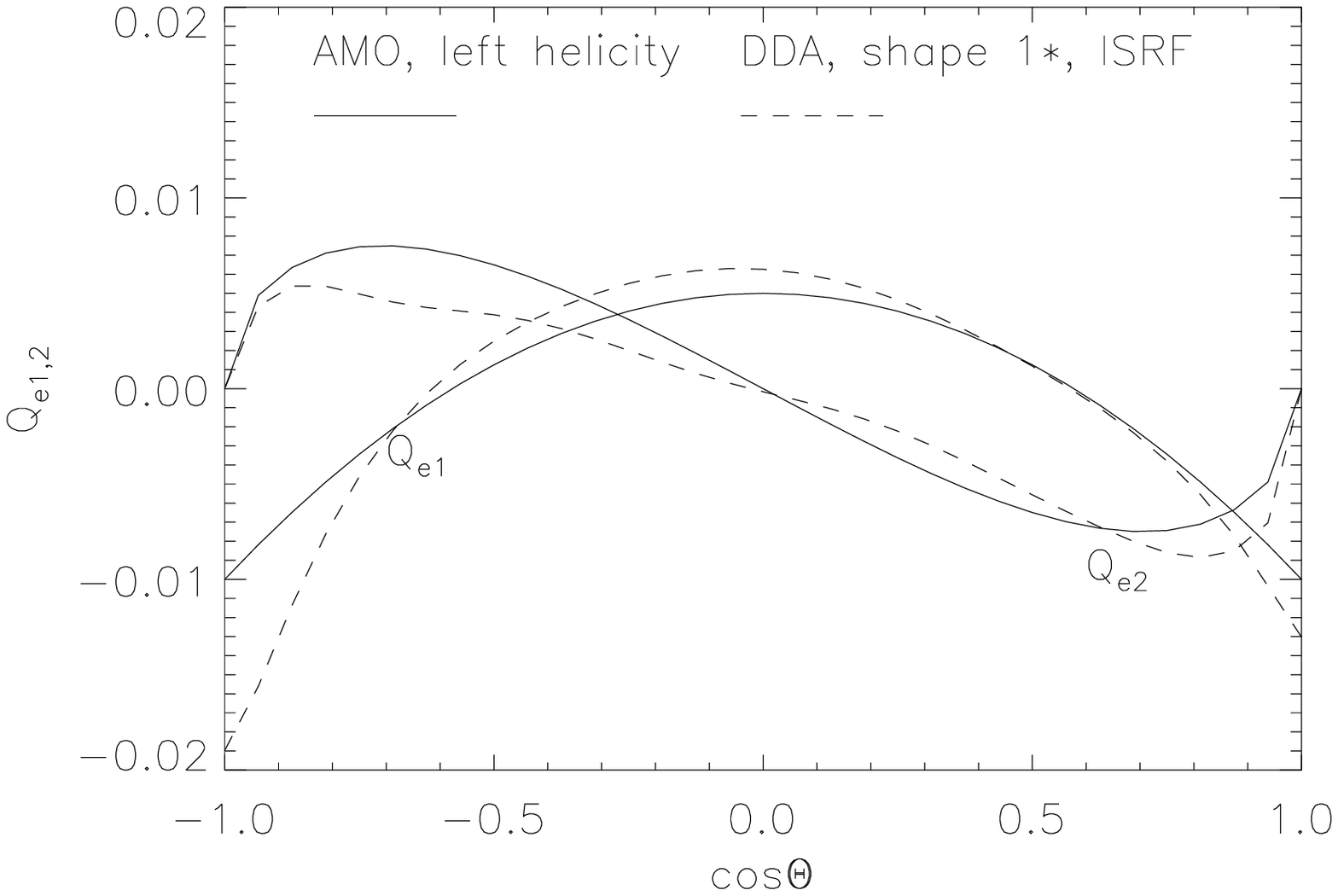}
\hfill
\includegraphics[width=3.in]{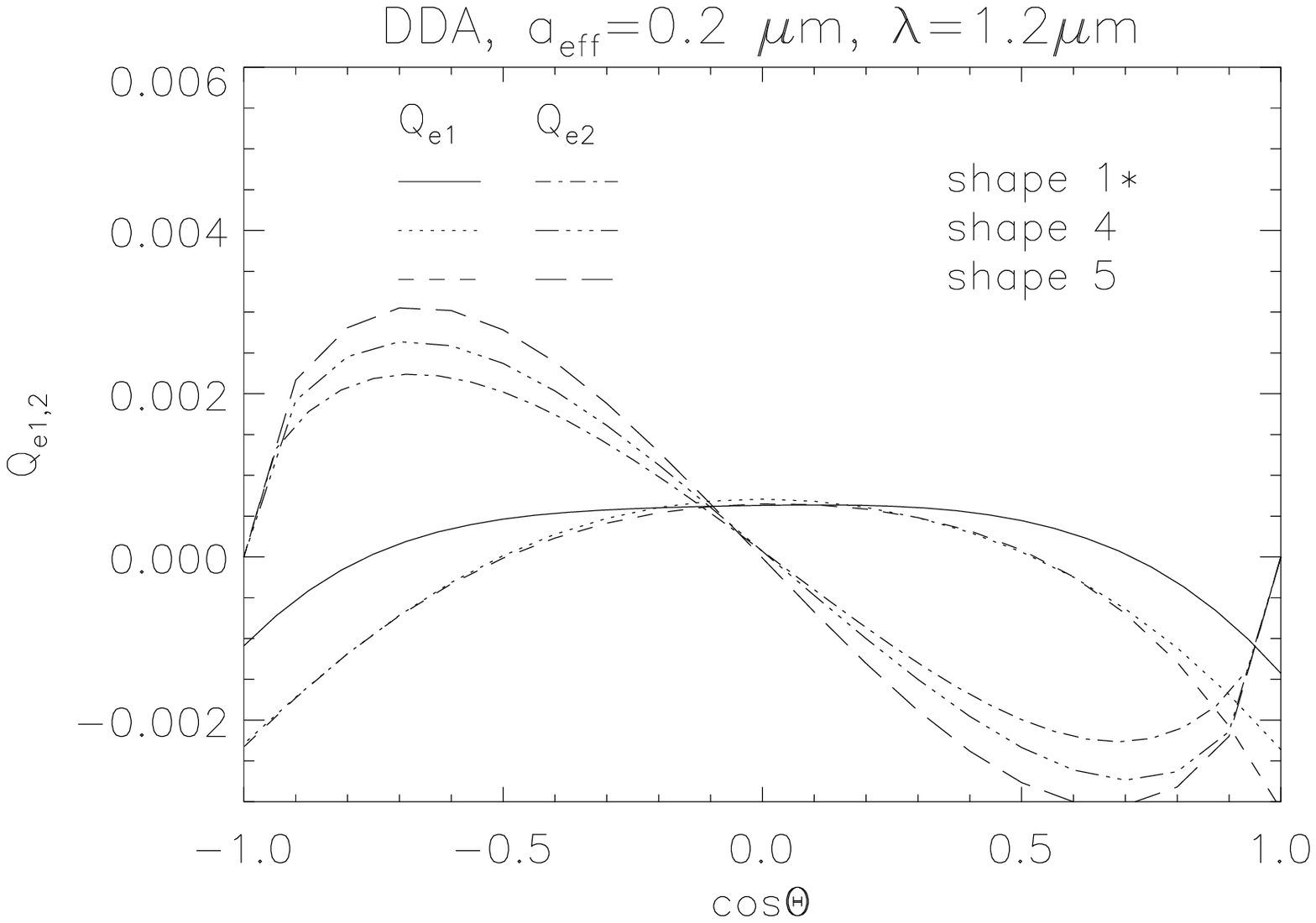}
\caption{
\small
{\it (a) Left panel}.-- Two components of the radiative torques
are shown for our
 analytical model (solid lines) in Fig.~7a and for an irregular
grain in Fig.~8 (dashed lines).
 {\it (b) Right panel}.-- Radiative torques for different grain shapes. From
Lazarian \& Hoang (2007a). }
\label{torques}
\end{figure}

\begin{figure}[h]
\includegraphics[width=2.8in,angle=270]{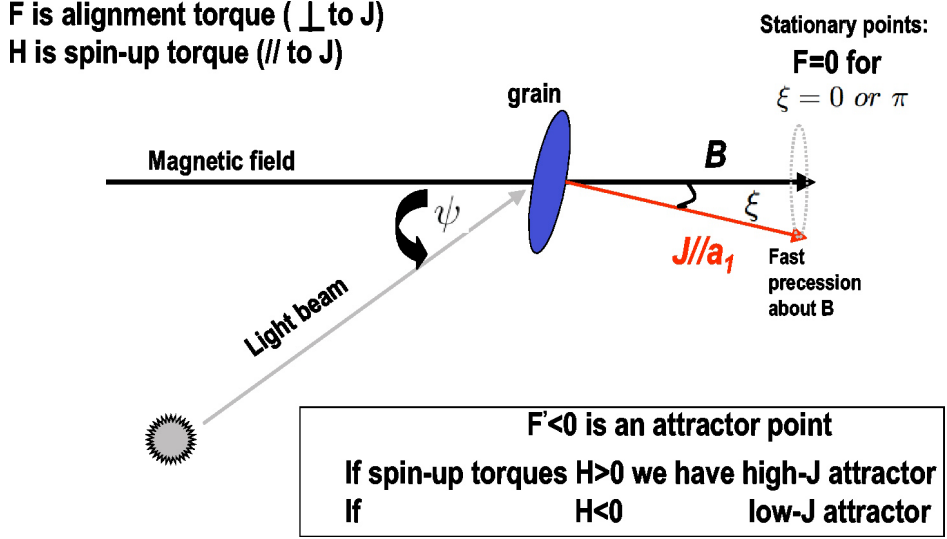}
\caption{A simplified explanation of the grain alignment by radiative torques.The grain, which is depicted as a spheroid in the figure, in fact, should be irregular to get non-zero radiative torque. 
The positions ${\bf J}$ parallel (or anti-parallel) to ${\bf B}$ correspond to the stationary points as at these positions the component of torques that changes the alignment angle vanishes. As 
internal relaxation makes ${\bf J}$ aligned with the axis $a_1$ of the maximal moment of grain inertia, the grain gets aligned with long axes perpendicular to ${\bf B}$.}
\label{alignment}
\end{figure}

The analytical model for radiative torque alignment was proposed in Lazarian \& Hoang (2007a). 
This simple model was shown to reproduces well the essential basic properties of radiative
torques. The model consists of an ellipsoidal grain with a mirror attached to
its side (see Fig.~\ref{AMO}). This model allows an analytical treatment and provides
a physical insight why irregular grains get aligned. Fig.~\ref{alignment} illustrates why radiative torques tend to align grains the "right way", i.e. in agreement with observations. Interstellar grains experience internal relaxation that tends to make them rotate about their axis of maximal moment of inertia. Therefore, it is sufficient to follow the dynamics of angular momentum to determine grain axes alignment. Let us call the component of torque parallel to ${\bf J}$ the {\it spin-up torque}
${\bf H}$ and perpendicular to ${\bf J}$ the {\it alignment torque} ${\bf F}$. The angular momentum ${\bf J}$ is precessing about magnetic field due to the magnetic moment of a grain (see
Dolginov \& Mytrophanov 1976). The alignment torques ${\bf F}$ are perpendicular to ${\bf J}$ and therefore as  ${\bf J}$ gets parallel to ${\bf B}$ the fast precession of the grain makes the torques averaged over ${\bf J}$ precession vanish as $\xi\rightarrow 0$. Thus the positions corresponding to ${\bf J}$ aligned with ${\bf B}$ are stationary points, irrespectively of the functional forms of radiative torques, i.e.
of components $Q_{e1}(\Theta)$ and $Q_{e2}(\Theta)$. In other words, a grain can stay aligned\footnote{The arguments above are quite general, but they do not address the question whether there are other stationary points, e.g. whether the alignment can also happen with ${\bf J}$ perpendicular to ${\bf B}$.
To answer this question one should use the actual expressions for  $Q_{e1}(\Theta)$ and $Q_{e2}(\Theta)$. The analysis in Lazarian \& Hoang (2007a) shows that there is, indeed, a range of angles
between the direction of radiation and the magnetic field for which grains tend to aligned in a "wrong" way, i.e. with long axes parallel to magnetic field. However, this range of angles is rather narrow and does not exceed several degrees and, as we discuss further, this range narrow that the range of grain thermal
wobbling.} with $\xi=0$ or $\pi$.

As ${\bf J}$, due to internal relaxation (see \S\ref{wobbling}) the
grains tend to rotate with ${\bf J}$ aligned with grain shortest axes (corresponding to the maximal moment of inertia), the
rotation with long grain axes perpendicular to magnetic field corresponds to the stationary point. If at the stationary point
the spin-up torques are positive, the grain rotates fast and we have a high-$J$ attractor point. If at the stationary point the
the torques spin the grain down, we have a point of slow rotation, or low-$J$ attractor point. Interestingly enough, the alignment
happens also in the absence of internal relaxation (see Hoang \& Lazarian 2008c).

\begin{figure}[h]
\includegraphics[width=3.in]{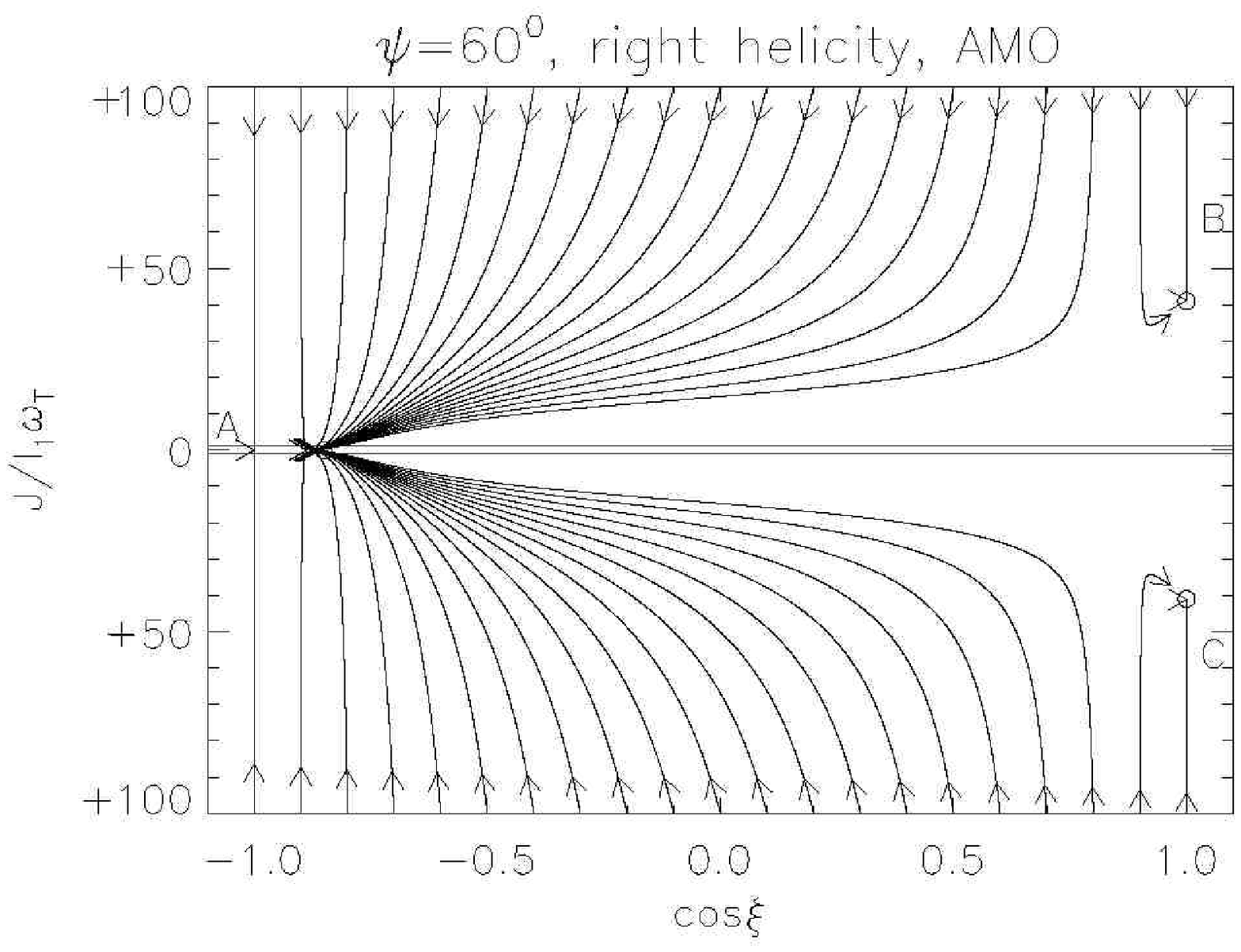}
\hfill
\includegraphics[width=3.in]{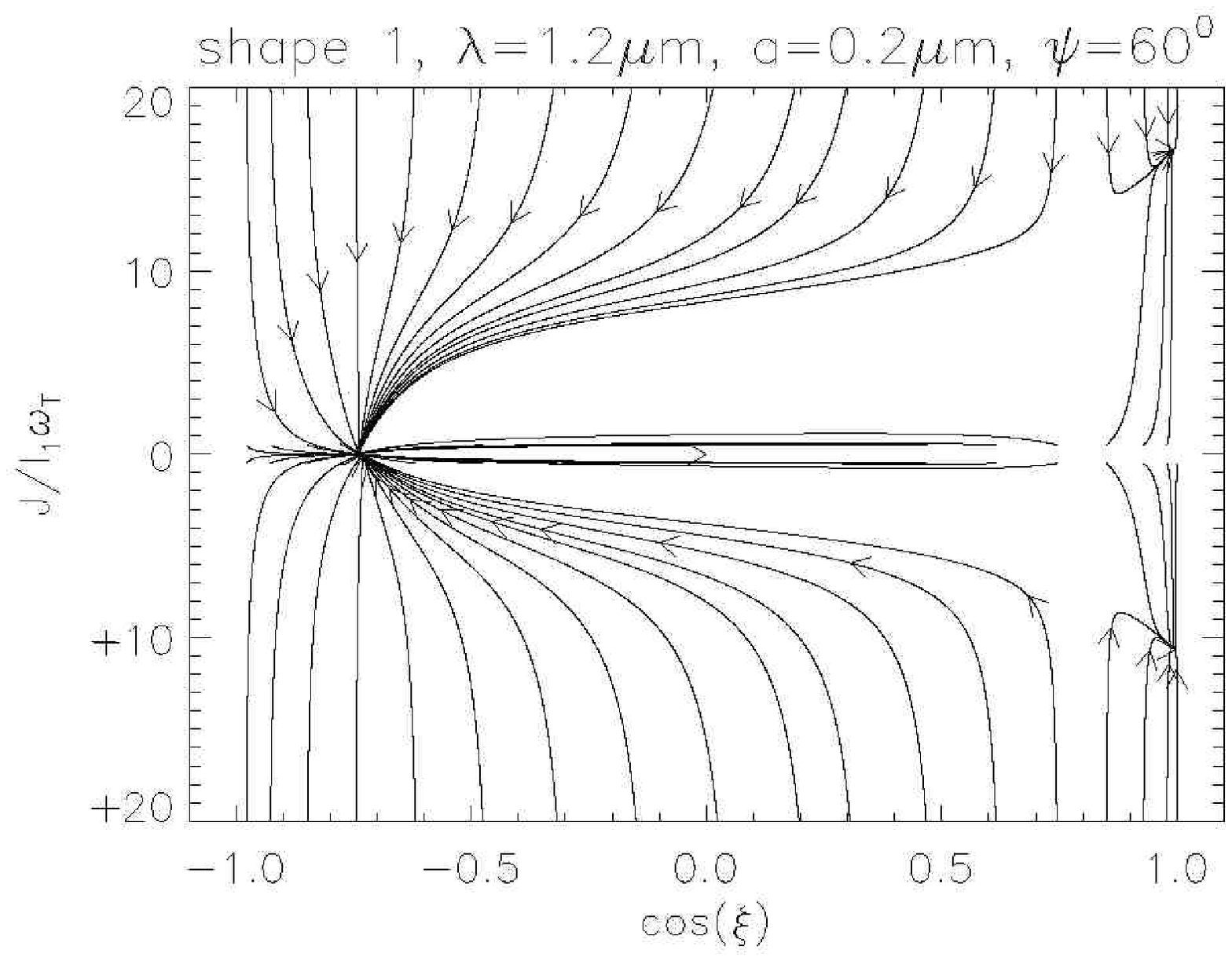}
\caption{\small Phase trajectory maps are obtained here by assuming that grains have initially some large values of angular momentum. The trajectories evolve
as a result of radiative torque action. {\it (a) Left panel}.-- Phase trajectory map obtained for the
model grain given shown in Fig.~\ref{AMO}.
 {\it (b) Right panel}.-- The same for an irregular grain in Fig.~\ref{shapes} (shape 1).
 From Lazarian
\& Hoang (2007a).}
\label{traject}
\end{figure}

 Note,  that grains can be both  
``left-handed'' and  "right-handed". The two species of different handedness exhibit their sets of
$Q_{e1}$ and $Q_{e2}$ functions that are described by the analytical model (see Lazarian \& Hoang 2007a).
For our grain model to become ``right handed'' the mirror should be turned  by 90 degrees. The introduction of the analytical model in Lazarian \& Hoang (2007a) ushered in the quantitative theoretical modeling of radiative torque alignment. However, to understand it one should recall the key features of grain dynamics related to grain internal relaxation. This is done in the next section. The visual comparison of
the radiative torques acting on arbitrarily shaped grains shown in Fig.~\ref{shapes} and the predictions for the torques arising from a model in Fig.~\ref{AMO} is given in Fig.~\ref{torques}. For this comparison we changed the handedness of the shape 1 grain (denoted therefore shape 1* in the Fig.~\ref{torques}). The phase trajectories of irregular grains and model grains are similar (see Fig.~\ref{traject}). 

\section{Grain dynamics: grains wobbling, flipping and thermally trapped}
\label{wobbling}
\subsection{Internal relaxation}
\label{w1}

 To produce the observed starlight polarization, grains must be aligned, with their
 {\it long axes} perpendicular to magnetic field. According to eq. (\ref{R}) this
 involves alignment not only of the grains' angular momenta ${\bf J}$ with respect to
 the external magnetic field ${\bf B}$, but also the alignment of the grains'
 long axes with respect to ${\bf J}$ (see Fig.~\ref{grain}). The latter alignment is frequently termed the "internal alignment".

Purcell (1979) was the first to consider internal relaxation as the course of
internal grain axes alignment. He showed that the rates of internal relaxation of energy in wobbling grains
are much faster than the time of grain alignment. This induced many researchers to think that
all grains rotate about the axis of their maximal inertia and the internal alignment is generically is perfect.
 Lazarian (1994) corrected
this conclusion by showing that the fluctuations associated with the internal dissipation through the Fluctuation Dissipation Theorem should induce the wobbling, which amplitude increases as the effective rotational temperature of grains approaches that of grain material.

The rate of internal relaxation is an important characteristic of grain dynamics. Spitzer \& McGlynn (1979), Lazarian \& Draine (1997, 1999ab), Lazarian \& Hoang (2008), Hoang \& Lazarian (2008b) demonstrated its crucial significance for various aspects of grain alignment. Purcell (1979) was the first to evaluate the rates of internal relaxation, taking into account the inelastic effects\footnote{A more recent study of inelastic relaxation is provided in Lazarian \& Efroimsky (1999).} and a particular effect that he discovered himself, the Barnett relaxation. As we know, the Barnett effect is the magnetization of a paramagnetic body as the result of its rotation. Purcell (1979) noticed that the wobbling rotating body would experience changes of the magnetization arising from the changes of the direction of rotation in grain axes. This would entail fast relaxation, with
a characteristic time about a year for a $10^{-5}$~cm grains. Lazarian \& Draine (1999a) identified $10^6$ times stronger relaxation related to nuclear spins and Lazarian \& Hoang (2008) showed that superparamagnetic grains, e.g. grains with  magnetic inclusions (see Jones \& Spitzer 1967, Mathis 1986, Martin 1994), exhibit the enhanced internal relaxation. 

\begin{figure}
\includegraphics[width=4.3in]{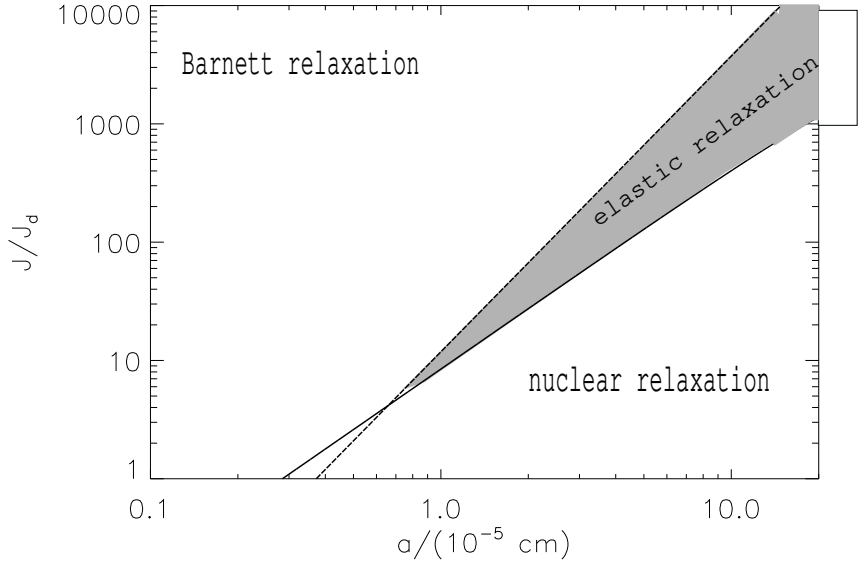}
\caption{ \small The relative role of the Barnett, nuclear and inelastic relaxation for grains of different sizes. Rates of Barnett relaxation are from Purcell (1979), nuclear relaxation are from Lazarian \& Draine (1999b), inelastic relaxation are from Lazarian \& Efroimsky (1999).}
\label{relax}
\end{figure}

The relative role of internal relaxation for the Barnett, nuclear and inelastic relaxation is shown in Fig.~\ref{relax}. The calculations of inelastic relaxation from Lazarian \& Efroimsky (1999) are used for the plot. 
These rates are important for grain dynamics. In particular, they influence how grain crossovers happen (see \S\ref{pinwheel}). 

\subsection{Regular pinwheel torques}
\label{regular}    

\begin{figure}
\includegraphics[width=2.3in]{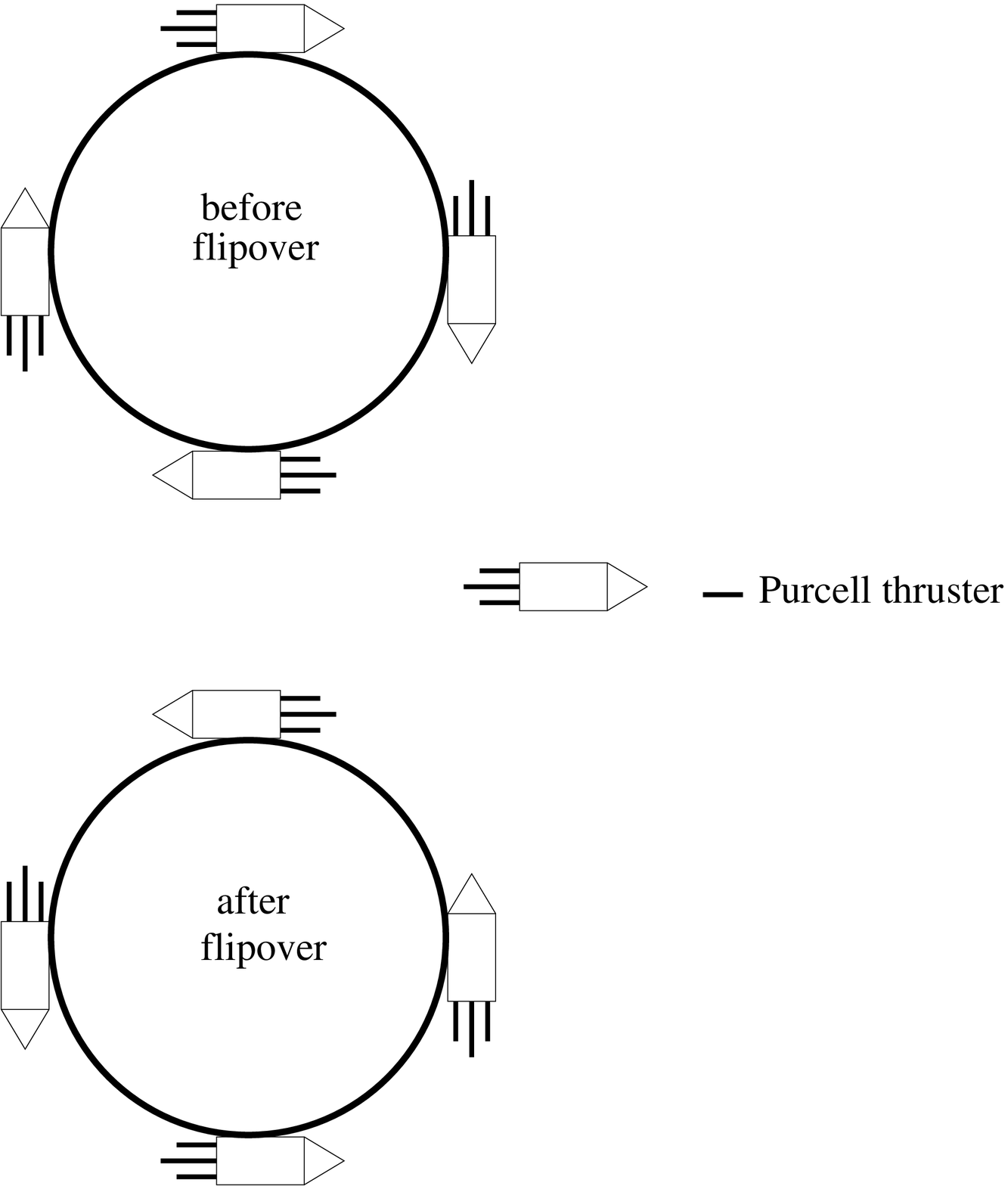}
\hfill
\includegraphics[width=2.0in]{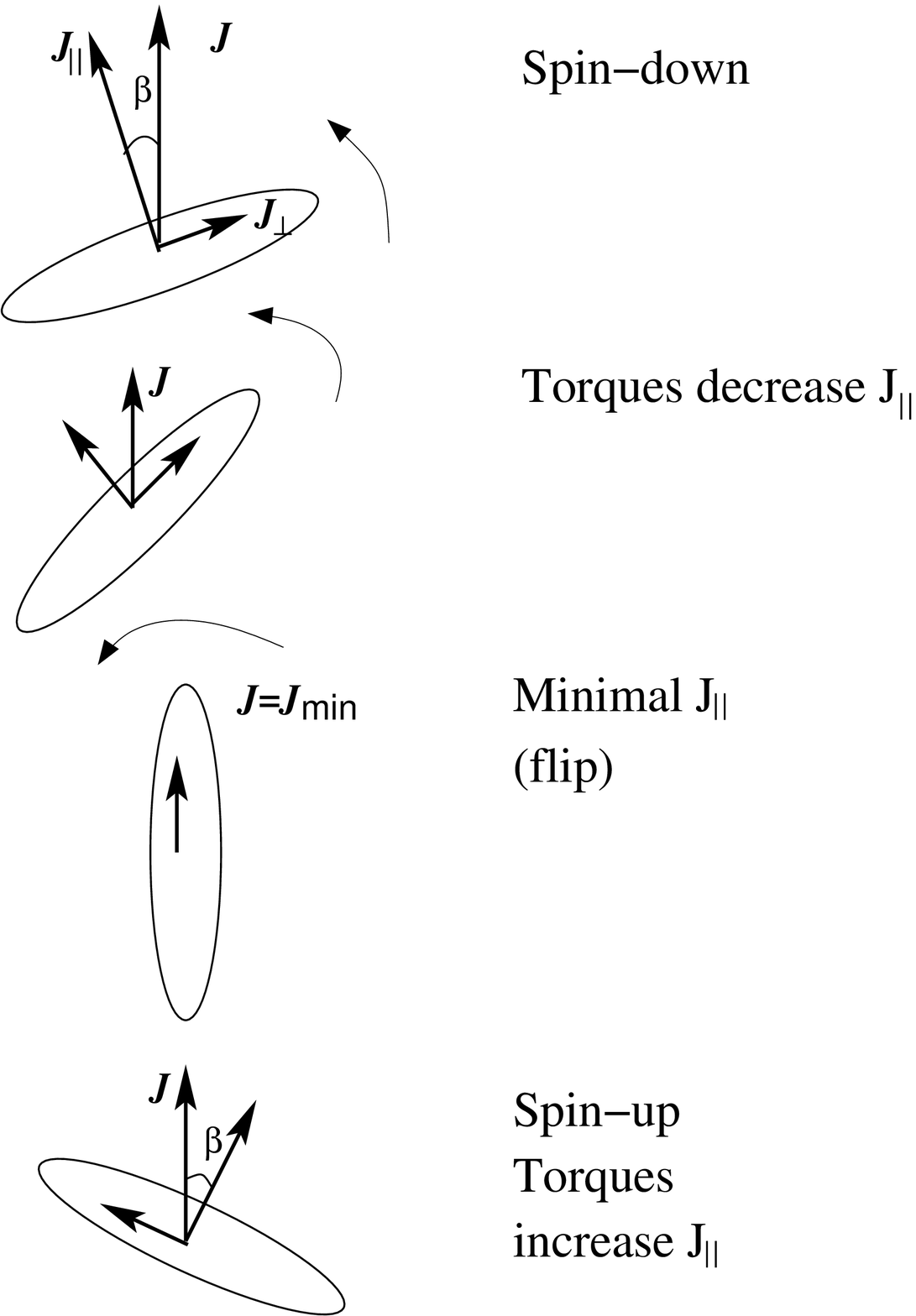}
\caption{ \small {\it (a) Left panel}-- A grain acted upon by Purcell's
torques before and after a flipover event. As the grain flips, the direction
of torques alters. The H$_2$ formation sites act as thrusters. {\it (b) Right
panel}-- A regular crossover event as described by Spitzer \& McGlynn (1979).
The systematic torques nullify the amplitude of the ${\bf J}$ component
parallel to the axis of maximal inertia, while preserving the other component,
$J_{\bot}$. If $J_{\bot}$ is small then the grain is susceptible to
randomization during crossovers. The direction of ${\bf J}$ is preserved in
the absence of random bombardment.}
\label{flips}
\end{figure}

 Purcell (1975, 1979) realized that
grains may rotate at a rate much faster that the thermal rate if they are subject to systematic torques. P79
 identified three separate systematic torque
mechanisms: inelastic scattering of impinging atoms when gas and grain
temperatures differ, photoelectric emission, and H$_2$ formation on grain
surfaces (see Fig.~\ref{flips}). Below we shall refer to the latter as "Purcell's
torques". These were shown to dominate the other two for typical conditions in
the diffuse ISM (P79).  The existence of systematic H$_2$ torques is expected
due to the random distribution over the grain surface of catalytic sites of
H$_2$ formation, since each active site acts as a minute thruster emitting
newly-formed H$_2$ molecules. A later study of uncompensated torques in
Hoang \& Lazarian (2008b) added additional systematic torques to the list, namely,
torques arising from plasma-grain interactions and torques arising from emission
of radiation by an irregular grains. Radiative torques arising from the interaction
of the {\it isotropic} radiation with an irregular grain (Draine \& Weingartner 1996) 
also represent the systematic torques fixed in the grain body. We shall call all the systematic
torques above {\it pinwheel torques} to distinguish them from the systematic torques arising 
from anisotropic flow of photons or atoms
interacting with an helical grain (see below).

Purcell (1979)  considered changes of the grain surface properties and noted that those
should stochastically change the direction (in body-coordinates) of the
systematic torques. Spitzer \& McGlynn (1979, henceforth SM79) developed a
theory of such {\it crossovers}. During a crossover, the grain slows down,
flips, and thereafter is accelerated again (see Fig.~\ref{flips}).

From the viewpoint of the grain-alignment theory, the important question is
whether or not a grain gets randomized during a crossover. If the value of the
angular momentum is small during the crossover, the grains are susceptible to
randomization arising from atomic bombardment. The original calculations in
Spitzer \& McGlynn (1979) obtained only marginal correlation between the values of the angular
momentum before and after a crossover, but their analysis disregarded thermal
fluctuations within the grain with temperature $T_{dust}$. According to the Fluctuation-Dissipation
Theorem these thermal fluctuation induce thermal wobbling with the frequency determined
by the internal relaxation rate (see Lazarian 1994). 
If the crossover happens over time larger than the grain internal relaxation time, Lazarian \& Draine (1997) showed
that the Spitzer \& McGlynn (1979)
theory of crossovers should be modified to include the value of thermal angular momentum
$J_{d, \bot}\approx (2kT_{dust} I_{\|})^{1/2}$, where $I_{\|}$ is the maximal moment of inertial
of an oblate grain. What is the size of grains for which the effect gets important? When nuclear
relaxation is accounted for, this provides a size of $a_c\approx 10^{-4}$~cm. Thus for large grains, 
e.g. in dark clouds and accretion disks the modification of the Spitzer \& McGlynn (1979) theory is important.

\begin{figure}
\includegraphics[width=4.3in]{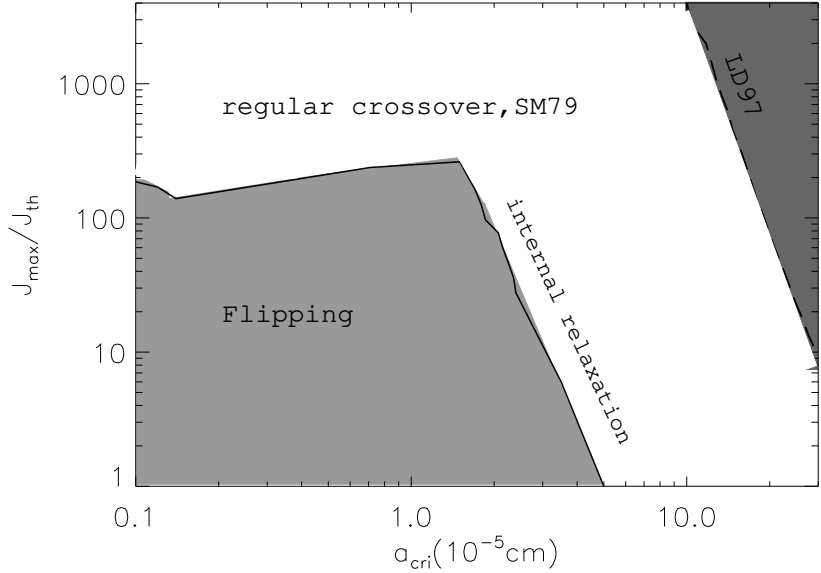}
\caption{ \small The applicability range for the Spitzer \& McGlynn (1979), Lazarian \& Draine (1997) and Lazarian \& Draine (1999ab) theories.
The latter predicts flipping the rate of which were elaborated in Hoang \& Lazarian (2008b). $J_{max}$ parameterizes here the value of the pinwheel torques,
i.e. $J_{max}= t_{gas} dJ/dt$, where the derivative $dJ/dt$ arises from the pinwheel torque action.}
\label{range}
\end{figure}

\subsection{Thermal trapping of grains}
\label{trapping}

What would happen for small grains? Lazarian \& Draine (1999a) predicted that
such grains would flip fast and this will alter the Purcell torques making grain
rotate thermally in spite of the presence of the uncompensated torques (see Fig.~\ref{flips}).  
This way Lazarian \& Draine (1999a) explained why interstellar grains smaller than $10^{-6}$~cm poorly aligned
via paramagnetic relaxation. A more elaborate study of the flipping phenomenon in Roberge \& Ford
(2000 preprint; see also Roberge 2004) supported this conclusion. Weingartner (2008) challenged this 
result by showing that spontaneous thermal flipping does not happen if the internal relaxation
diffusion coefficient (Lazarian \& Roberge 1997) is obtained with a different integration constant. Does this
invalidate the phenomenon of thermal trapping predicted in Lazarian \& Draine (1999a)? Our study in
Hoang \& Lazarian (2008b) shows that the flipping and trapping do happen, if one takes into account additional
fluctuations associated, for instance, with the action of the uncompensated torques. 

What does happen with the grains which do not flip, but smaller than $a_c$? For those grains the original
SM79 theory of crossovers is applicable and these grains are marginally aligned via Davis-Greenstein process, provided
that the pinwheel torques are short-lived, i.e. the time scale of their existence is much smaller than the time
of paramagnetic relaxation. Such grains rotate at high rate in accordance with the Purcell (1979) model.  As we discuss
further, the fact that pinwheel torques are not suppressed by flipping may allow better alignment by radiative torques. 

Fig.~\ref{range} defines the range over which the different model of crossovers are applicable. It is clear that for a sufficiently
large $J_{max}$ the flipping gets suppressed.

\section{Quantitative modeling of radiative torques}

\subsection{Predictions of the analytical model}
\label{predictions}

How can one characterize radiative torques acting on grains of {\it arbitrary} shapes? 
Our work showed that the entire description of alignment may be obtained with
the two components of the radiative torques $Q_{e1}$ and $Q_{e2}$ as they
are defined in the caption of Fig.~\ref{AMO}. The third component $Q_{e3}$ is
responsible for grain precession only. 
The functional dependences of the torque components
 that are experienced by our model grain are similar to those
experienced by irregular grains shown in Fig.~\ref{shapes}. It is really remarkable that
our model and grains of very different shapes have very similar
functional dependences of their torque components (see Fig.~\ref{torques})!

\begin{figure}
\includegraphics[width=5.3in]{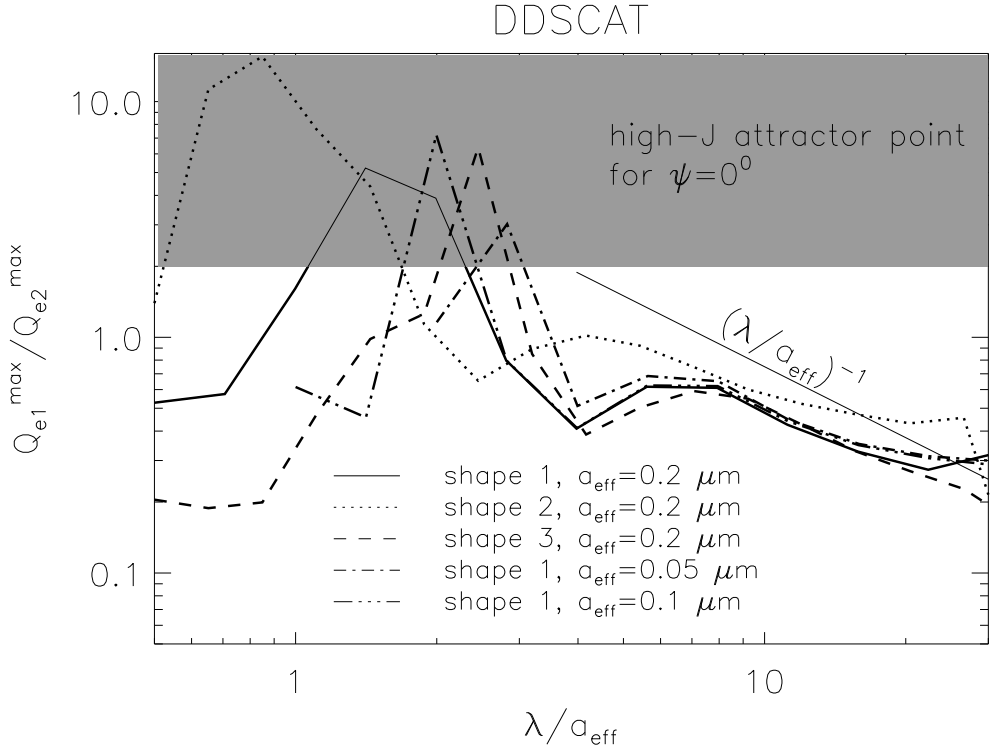}
\caption{\small The magnitude of the ratio $q=Q_{e1}^{max}/Q_{e2}^{max}$ that characterize the radiative torque alignment of grains depends on both grain shape and the wavelength of radiation. From
Lazarian \& Hoang 2007a.}
\label{dependence}
\end{figure} 

\begin{figure}
\includegraphics[width=5.3in]{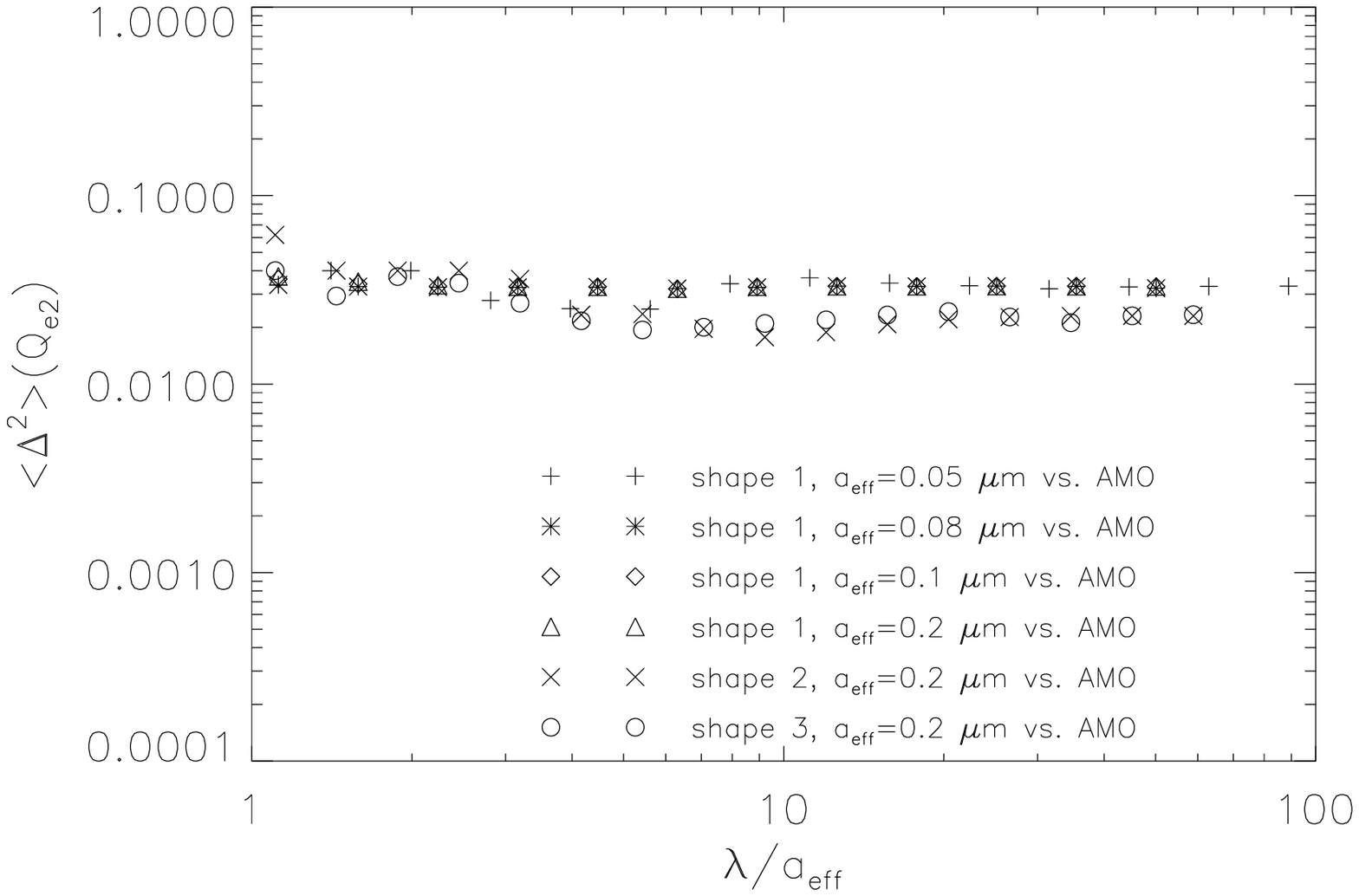}
\caption{\small Numerical comparison of the torques calculated with DDSCAT for irregular grains for different wavelength and the analytical model (AMO) of a helical grain (from Lazarian \& Hoang 2007a).}
\label{chi}
\end{figure}

For typical conditions of the diffuse ISM the radiative torques are strong and
absolutely dominate the grain dynamics. In this situation, the alignment depends
only on the ratio of the two components of torques, namely, on $Q_{e1}/Q_{e_2}$.
This ratio is a function of the angle between the radiation direction, but ,as the functional
dependences of $Q_{e1}$ and $Q_{e2}$ for irregular grains correspond well to the
analytical expressions obtained for a helical model grain in one can clearly see from Fig~\ref{chi}, 
where the function
\begin{equation}
\langle \Delta^2\rangle(Q_{e2})=\frac{1}{\pi (Q_{e2}^{max})^2} \int^{\pi}_{0} \left[Q_{e2}^{irregular}(\Theta) -Q_{e2}^{model} (\Theta)\right]^2 d\Theta
\label{chi_eq}
\end{equation}
characterizes the deviation of the calculated torques $Q_{e2}$ calculated numerically for irregular grains from the analytical prediction of the torques in Lazarian \& Hoang (2007a) model.  

It is very important that one can fully characterize
the ratio of these two torques by defining the ratio of their amplitude values. In other words,
the knowledge of the ratio $q=Q_{e1}^{max}/Q_{e2}^{max}$ is sufficient to characterize the 
alignment by radiative torques. This is the second remarkable fact on the properties of
radiative torques. In terms of practical calculations, this enormously simplifies the calculations
of radiative torques: instead of calculating two {\it functions} $Q_{e1}(\Theta)$ and $Q_{e2}(\Theta)$ it
is enough to calculate just two {\it values} $Q_{e1}^{max}$ and $Q_{e2}^{max}$. According
to Lazarian \& Hoang (2007a) the maximal value of the function $Q_{e1}(\Theta)$ is achieved for $\Theta=0$ of the function $Q_{e2}(\Theta)$ is
achieved at $\Theta=\pi/4$. In other words, one can use a {\it single ratio} $q=Q_{e1}^{max}/Q_{e2}^{max}=Q_{e1}(0)/Q_{e2}(\pi/4)$ instead of
{\it two functions} to characterize grain alignment. Thus, it is possible to claim that the $q$-ratio is as important for the alignment as the grain axis
ratio is important for producing polarized radiation by aligned grains. 

What does affect the $q$-ratio? One can see that it is not universal. The geometric optics calculations
for the model grain in Fig.~\ref{AMO} show that this ratio varies with the change of the angle that
characterizes the twist of the mirror in respect to the axis of inertia of the grain\footnote{This change somewhat
alters the functional dependences of the radiative torques  and we find that the best agreement can be obtained
by fixing the angle of the mirror twist at $\pi/4$.} Similarly, we can see in Fig.~\ref{dependence} that the $q$-ratio is a function
of both the grain shape and radiation wavelength.  A general tendency is that $q$ peaks when grains are of
the size comparable to the radiation wavelength and decreases below unity when grains are substantially smaller
than the radiation that they interact. 

\begin{figure}
\includegraphics[width=5.3in]{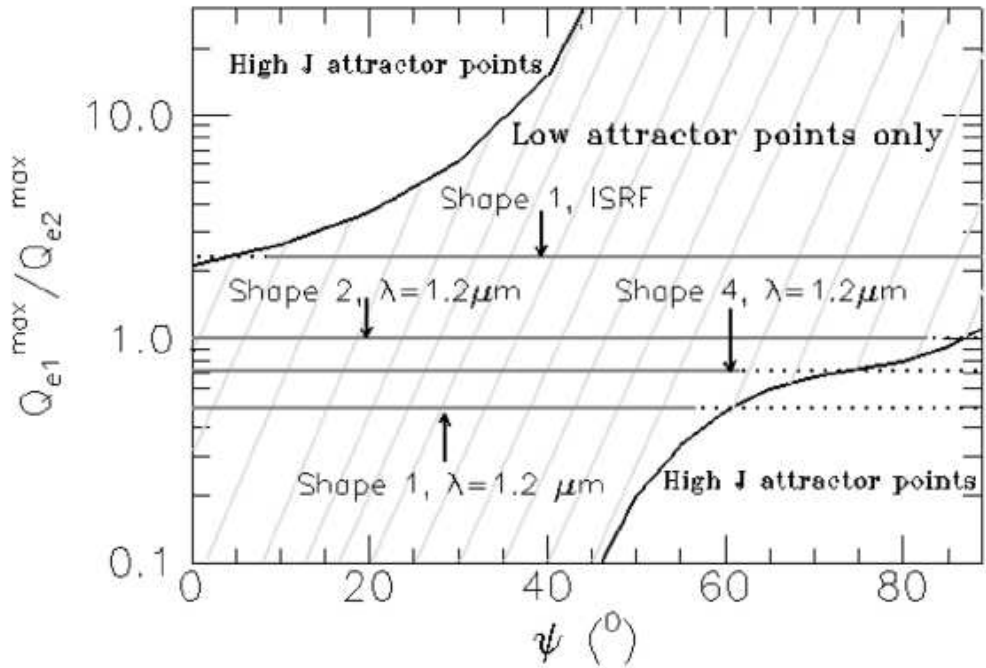}
\caption{\small Parameter space for which grains have only low-$J$ attractor point and both low-$J$ and high attractor point (from Lazarian \& Hoang 2007a). In the situation when the high-$J$ attractor point
is present grains eventually get there and demonstrate perfect alignment. In the situation when only low-$J$ attractor point is present, the alignment is partial.}
\label{parameter}
\end{figure} 

The practical use of the $q$-ratio is presented in Fig.~\ref{parameter} where the parameter space for the existence of
only low-$J$ as well as coexistence of both high-$J$ and low-$J$ attractor points is shown. As we discussed
earlier, in the case when high-$J$ attractor point exists, grains eventually get into that position, which is
characterized by both the fast rotation and {\it perfect} alignment in respect to magnetic field. In the situation when
only low-$J$ attractor point exists, the alignment in respect to magnetic field is not stable and the characteristic degrees
of alignment vary on average from 20\% to 50\%. Comparing results in Fig.~\ref{dependence} and \ref{parameter} one may conclude
that in the situations when the alignment arises from the radiation with the wavelength $\lambda$ larger than the grain
characteristic size, the perfect alignment is  likely when the angle $\psi$ between the radiation direction and magnetic field
is larger than $\pi/4$.

 Are other alignment directions feasible? Due to the fast Larmor precession of grains, the magnetic field constitute for them the symmetry axis and therefore the alignment can happen either with long axes perpendicular to magnetic field ("right alignment) or with long axes parallel to magnetic field ("wrong alignment"). 
The model in Lazarian \& Hoang (2007a) shows that for a range of angles between the radiation and the magnetic
field the alignment gets ``wrong'', i.e. with the long axes parallel to magnetic field. However, this range is narrow (limited
to radiation direction nearly perpendicular to magnetic field). Therefore the range of "wrong" alignment disappears in the
presence of grain thermal wobbling described in \S\ref{wobbling} as was shown in Hoang \& Lazarian (2008a). 

All the calculations above assumed that the radiation comes from a single direction. Is it possible to
provide prescriptions for the alignment when the radiation field is characterized by smooth distribution of
light coming from different directions. In this case it is natural to decompose the
radiation field into multipoles and consider the alignment by individual components. This was done in Hoang \& Lazarian (2008b)  who considered the radiative torque alignment by the dipole and quadropole components of the radiation field. 

\subsection{Radiative torque alignment of superparamagnetic grains}
\label{magnetic}

Superparamagnetic grains, i.e. grains with enhanced paramagnetic relaxation, 
 were invoked by Jones \& Spitzer (1967) within the model of paramagnetic alignment. 
What does happen when the dynamics of grains is determined by radiative torques? We see from Fig.~\ref{parameter} that 
for a substantial part of the parameter space grains are driven to the low-$J$ states, i.e. {\it subthermally}, which is in contrast to
the assumption in Draine \& Weingartner (1996) that in the presence of radiative torques
most of the interstellar grains should rotate at  $T_{rot}\gg T_{gas}$.

\begin{figure}
\includegraphics[width=5.3in]{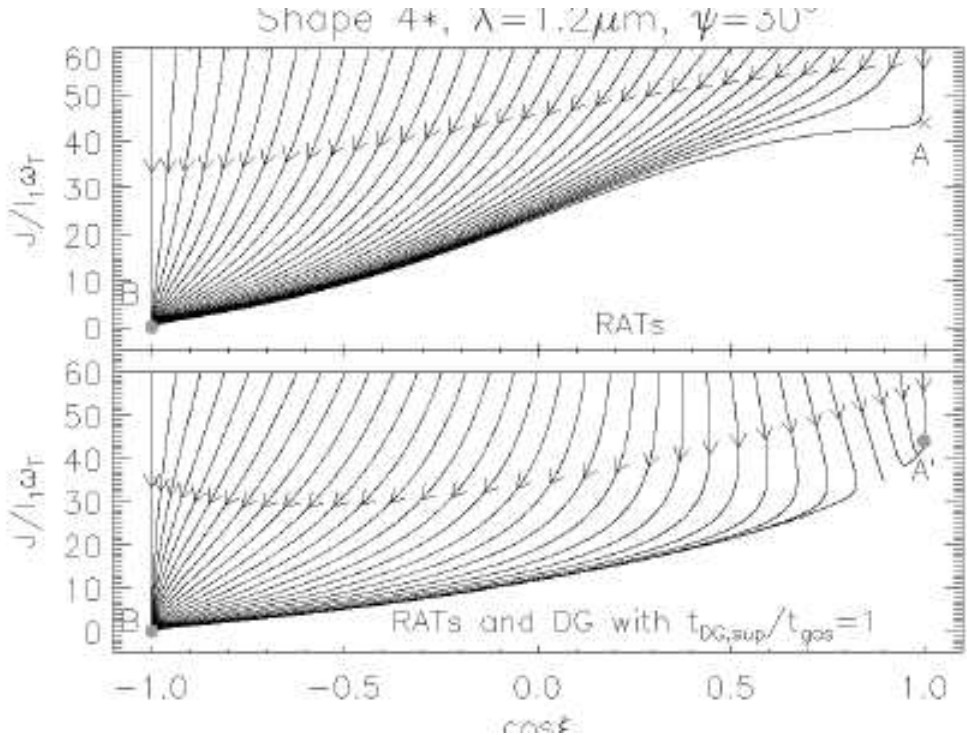}
\caption{\small A paramagnetic grain gets only a low-$J$ attractor point. For the same set of parameters a superparamagnetic grain gets also a high-$J$ attractor point.
The fact that most of the phase trajectories go in the direction of the low-$J$ attractor point illustrates the dominance of the radiative torques for the alignment even
in the case of the superparamagnetic grain. However, high-$J$ attractor points are more stable than the low-$J$ attractor points. As a result, all grains eventually end up
at the high-$J$ attractor point.}
\label{superparamagnetic}
\end{figure} 

The picture above, however, is different if grains paramagnetically dissipate the energy on the time scales shorter
than the gaseous damping time.  Lazarian \& Hoang (2008)  found that if grains have superparamagnetic inclusions, they {\it always} get 
high-$J$ attractor point. Fig.~\ref{superparamagnetic} shows that for superparamagnetic grains subject to a diffuse interstellar radiation field most grains still get
to the low attractor point, which reflects the fact that it is the radiative torques rather than paramagnetic ones that dominate the
alignment. As the high-$J$ attractor point is more stable compared to the low-$J$ attractor point, similar to the ordinary paramagnetic  grains,
superparamagnetic grains get transfered by gaseous collisions from the low-$J$ to high-$J$ attractor points. Thus, superparamagnetic grains 
always rotate at high rate in the presence of radiative torques. One concludes that, rather
unexpectedly, intensive paramagnetic relaxation changes the rotational state of the grains, enabling them to
rotate {\it rapidly}. The alignment of grains at high-$J$ point is {\it perfect}.

\subsection{Radiative torque alignment in the presence of pinwheel torques}
\label{pinwheel}

Pinwheel torques were considered by Purcell (1979) in the context of paramagnetic alignment.
How do these torques also affect the radiative torque alignment? Hoang \& Lazarian (2008b) showed
that the sufficiently strong pinwheel torques can create new high-$J$ attractor points. Therefore for
sufficiently strong pinwheel torques, e.g. for torques arising from H$_2$ formation, one may observe the correlation
of higher degree of polarization with the atomic hydrogen content in the media, provided that H$_2$ torques as strong
as they considered in Purcell (1979) and the subsequent papers (see Spitzer \& McGlynn 1979, Lazarian 1995, Lazarian \& Draine 1997). 
Another implicit assumption for observing this correlation is that the grains are not superparamagnetic. For superparamagnetic grains
the alignment, as we discussed above, is perfect anyhow. 

The existing uncertainty is related to the magnitude of the pinwheel torques. If the correlation is observed, this would mean that, first of all,
grains are not superparamagnetic and, in addition, the pinwheel torques are sufficiently strong to overwhelm the radiative torques which drive
the grains to  low attractor points. 

\begin{figure}
\includegraphics[width=2.9in]{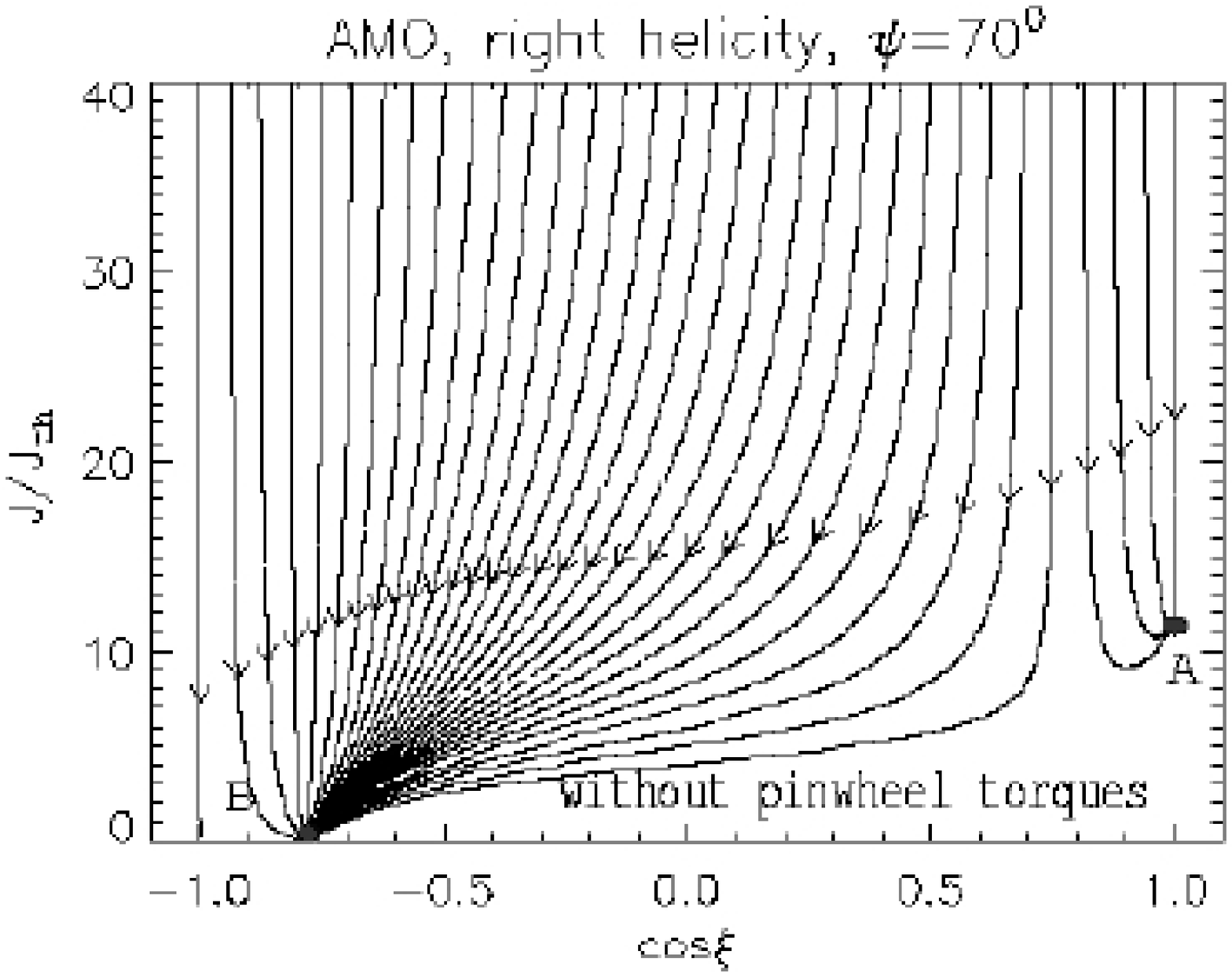}
\hfill
\includegraphics[width=2.7in]{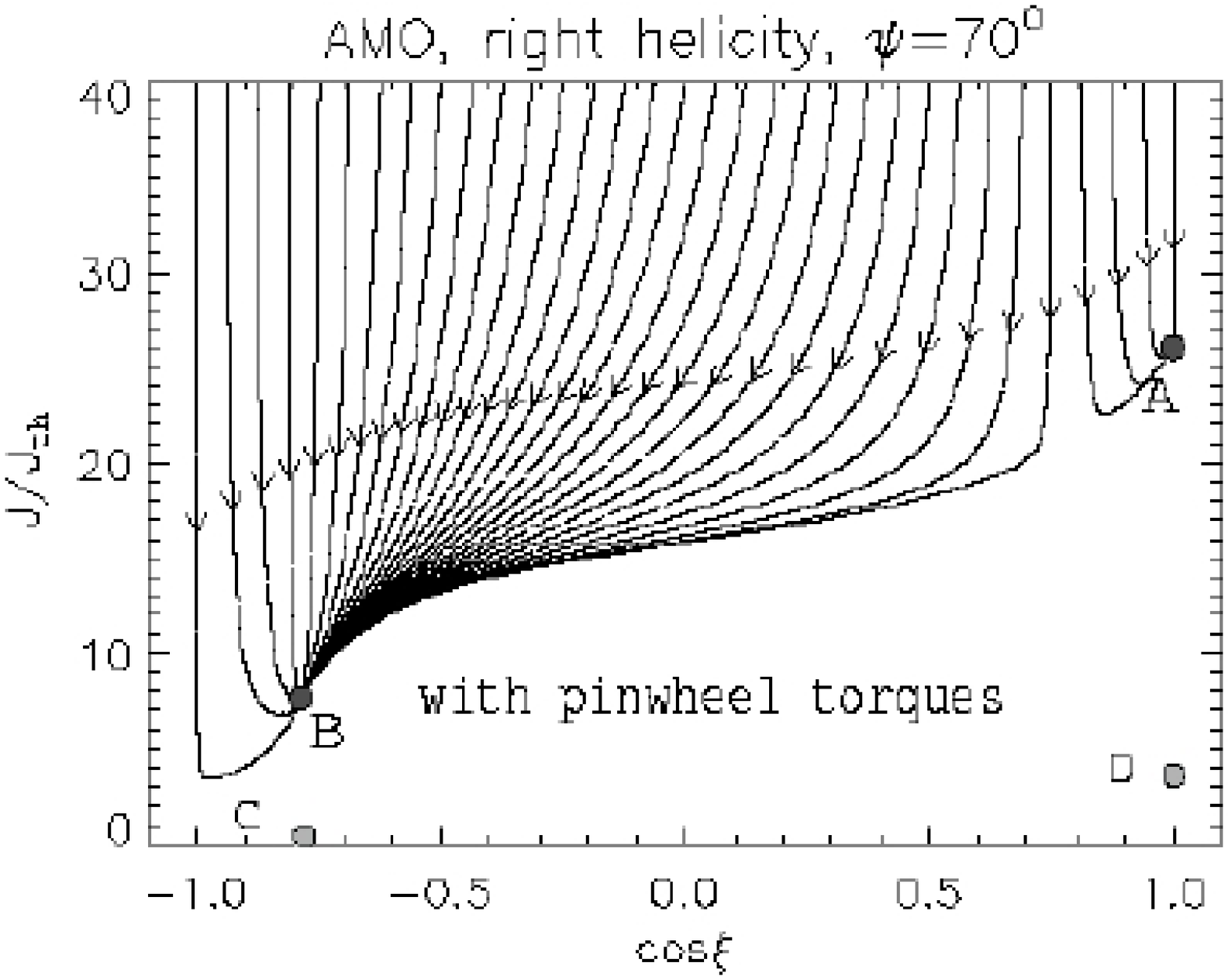}
\caption{\small  {\it Left panel}. Grain alignment by radiative torques in the absence of pinwheel torques. {\it Right panel}.  Grain alignment by radiative torques in the presence
of  pinwheel torques. The shown case corresponds to the presence of both the low-$J$ and high-$J$ attractor points in the absence of pinwheel torques. In the case when only a low-$J$ attractor
point exists the strong pinwheel torques lift the low-$J$ attractor point enhancing the alignment.}
\label{parameter}
\end{figure}

\section{Prospects of Grain Alignment Theory}

\subsection{Alignment of grains $>3\times 10^{-6}$~cm}

Grain alignment theory so far was mostly aimed at understanding of the alignment of large grains, e.g.
larger than $3\times 10^{-6}$~cm, for which we have the observational evidence of the alignment. Thus we,
first of all, deal with such grains.

Grain alignment theory has witnessed a substantial progress in the recent years. For the
first time in the history of the subject we can talk about the qualitative correspondence of the theoretical
predictions and observations (see Whittet et al. 2008). Other attempts of polarization modeling were made while
the quantitative understanding of radiative torques was only at its initial stage, but they also provide a satisfactory correspondence
with observations (see Cho \& Lazarian 2005, Pelkonen et al. 2007, Bethell et al. 2007). All this testify in
favor of radiative torques as a  dominant mechanism for the alignment of astrophysical grains. 

The questions that require further studies include resolving the issue of whether interstellar grains are superparamagnetic
or paramagnetic. As we discussed above, if grains are superparamagnetic, their alignment is much easier to predict. In most
of the interstellar conditions we expect such grains to be perfectly aligned with long axes perpendicular to the magnetic field.
If, however, grains are paramagnetic, they can be sometimes perfectly, if their $q$-ratio and the angle of illumination  corresponds
to the part of the parameter space in Fig.~\ref{parameter} that exhibit a high-$J$ attractor point. Otherwise, a reduced degree of
alignment is expected. 

How can one distinguish between the two cases above? It looks that two approaches are readily available. One is  based on statistical studies using
a simulated distribution of grains with different $q$ parameters and comparing the observations and simulations. The other  approach is
based on identifying of the situations when the geometry of illumination is known, as may be the case for some circumstellar regions, and
studying the alignment there. 

Alternatively, if the correlation of the degree of alignment with the pinwheel torques is detected for diffuse interstellar medium (see \S\ref{pinwheel}) this would also be the evidence that the grains are paramagnetic. Indeed, the interstellar radiation field is sufficiently strong render strong radiative torques essentially everywhere in the diffuse galactic gas. Therefore, the superparamagnetic grains are expected to be aligned perfectly according to \S\ref{magnetic} The variations, on the contrary, mean that the grain alignment is incomplete.

It is also important to identify the possible niches for other alignment processes, in particular to the mechanical alignment of helical grains. 
As we discussed earlier, the model grain in Fig.~\ref{AMO} is helical not
only in respect to radiation, but also to mechanical flows (see also LH07).
 It is easy to see that the
functional dependence of the torques does
not depend on whether photons or atoms are reflected from the mirror. Therefore
our extended discussion of radiative torque alignment above is applicable to the
mechanical alignment of helical grains. In the presence of 
dynamically important magnetic field, the alignment is
expected with long axes perpendicular to ${\bf B}$. If atoms stick to the
grain surface and then are ejected from the place of their impact, this changes the values
of torques by a factor of order unity. The helium atoms are likely to have specular impacts, however.

What would it take to make a grain helical for mechanical interactions? This
is a question similar to
one that worried researchers with the radiative torques before
Bruce Draine modified his code to treat radiative torques (see Draine 1996). We do not have the simulations of mechanical
torques on irregular grains, but in analogy with the radiative torques, one could
claim that such torques should be generic for an irregular grain. Detailed calculations
should clarify the actual degree of helicity of irregular grains and the magnitude of
the mechanical torques on such grains. The corresponding calculations are straightforward
and we expect to have the answer before the CMBPol launch.  

\subsection{Alignment of Small Grains}

Radiative torques predict the alignment of interstellar grains larger than $\approx 3\times 10^{-6}$~cm,
which is in an excellent agreement with the empirical Serkowski (1973) curve (see Lazarian 2007). 
For particles much less than the wavelength the efficiency of radiative
torques drops as $(a/\lambda)^4$ (see Lazarian 1995).
 Within circumstellar regions, where UV flux is enhanced
smaller
grains can be aligned by radiative torques. This could present a possible solution
for the reported anomalies of polarization in the 2175 {\,{\rm \AA}}
 ~~extinction feature (see Anderson et al 1996) which have been interpreted
as evidence of graphite grain alignment (Wolff et al 1997). If
this alignment happens in the vicinity of particular
stars with enhanced UV flux
and having graphite grains in their circumstellar regions, this may
explain why no similar effect is observed along other lines of sight.

The maximum entropy inversion technique in Kim \& Martin (1995) indicates that
 grains larger than a particular critical size are aligned. This is
consistent with our earlier discussion of radiative torques and the Serkowski law (see \S 6.2).
 However, an interesting feature of
the inversion is that it is suggestive of smaller grains being partially
aligned. Initially, this effect was attributed to the problems with the
assumed dielectric constants employed in the inversion, but a further
analysis that we undertook with Peter Martin indicated that the alignment
of small grains is real. Indeed, paramagnetic (DG) alignment must act on
the small grains\footnote{To avoid a confusion we should specify that we
are talking about grains of $10^{-6}$~cm. For those grains the results
of Davis-Greenstein relaxation coincide with those through resonance relation in
Lazarian \& Draine (2000). It is for grains of the size less than
$10^{-7}$~cm that the resonance relaxation is dominant.}.
An important feature of this weak alignment is that it is proportional to
the energy density of magnetic field. This opens a way for
a new type of magnetic field diagnostics.
As very small grains may emit polarized radiation as they rotate (see \S 4.2) both
UV and microwave polarimetry may be used to estimate the intensities of magnetic field.

\subsection{Alignment of PAHs}

In the range of frequencies from 30GHz to 90GHz the polarized emission from PAHs may play an important role. There have been
no systematic work on the PAHs alignment however. The fact that no polarization from infrared PAH emission has been detected
does not mean that the PAHs are not aligned. Indeed, the infrared emission arises from PAH absorption of a UV photon. The likely 
effect of such an absorption is the randomization of grain axes in terms of the angular momentum ${\bf J}$. This effect arises from the
coupling of rotational and vibrational degrees of freedom via internal relaxation (Lazarian 1994, Lazarian \& Roberge 1997, Weingartner 2008,
Hoang \& Lazarian 2008b). Therefore while ${\bf J}$ can be aligned with magnetic field, the grain axes may be randomized in respect to ${\bf J}$
during the short periods of PAH emission. On the contrary, the dipole emission arising from PAHs in accordance with the "spinning dust" process
in Draine \& Lazarian (1998) proceeds all the time and, moreover, is relatively insensitive to the randomization of grain axes. 

\begin{figure}
\includegraphics[width=4.3in]{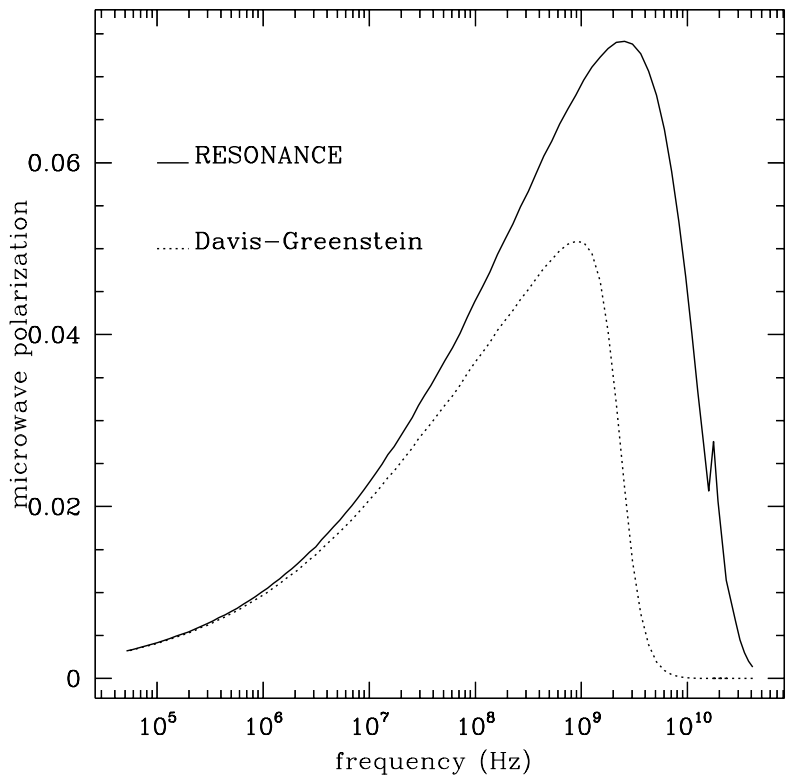}
\caption{\small Polarization for both
        resonance relaxation and Davis-Greenstein relaxation for grains in
        the cold interstellar medium as a function of frequency. From Lazarian \& Draine 2000.
        }
\label{microPAH}
\end{figure}

The only quantitative study of PAH alignment is one in Lazarian \& Draine (2000). There the so-called "resonance paramagnetic relaxation" process was appealed to. The traditional Davis-Greenstein (1951) relaxation is not applicable to small grains, as it disregards the important magnetization arising from the Barnett effect. This magnetization is very strong for the rapidly rotating PAHs. Moreover, the magnetization that arises is exactly of the right magnitude to induce paramagnetic resonance and therefore the dissipation much stronger than in the Davis-Greenstein process. Nevertheless, the polarization arising from the  PAHs predicted in Lazarian \& Draine (2000) is marginal for the emission at frequencies higher than 30GHz (see Fig.~\ref{microPAH})

\section{Grain alignment theory and CMBPol}

\subsection{Zodiacal dust as a polarized foreground}

The contribution from the Zodiacal dust is taken into account for the emission studies of CMB. At the same time,
radiative torque mechanism should make the Zodiacal dust at least partially aligned. Indeed, our study in 
Lazarian \& Hoang (2007a) shows that the radiative torques do not decrease when grain sizes are larger than the 
wavelength of impinging radiation, contrary to the predictions in Dolginov \& Mytrophanov (1976). Moreover,
Hoang \& Lazarian (2008c) showed that grains should be aligned even when they are so large enough that the
internal relaxation becomes negligible. The polarized emission by Zodiacal light may influence the studies by CMBPol
and therefore should be accounted for.

A possible study of the polarization arising from the Zodiacal light may be attempted in far infrared prior to the CMBPol mission.
However, even optical studies may clarify this issue, if we take into account that scattering from the aligned grains can result in
circularly polarized radiation, as it discussed in \S\ref{circular}

\subsection{"Magnetic" dust and "spinning dust" as polarized foregrounds}

The problem discussed in \S\ref{magnetic} was the superparamagnetic nature of the dust. Superparamagnetic and, in general,
strongly magnetic dust produces polarized microwave magneto-dipole emission, which can interfere with the CMB polarization 
studies (Draine \& Lazarian 1999). If the magnetic properties of dust are better constrained as a result of testing of grain
alignment predictions, this should simplify the removal of dust magnetic contribution at the microwave range.
Similarly, a better understanding of PAHs alignment should facilitate the removal of the "spinning dust" polarized contribution.

\begin{figure}
\includegraphics[width=4.3in]{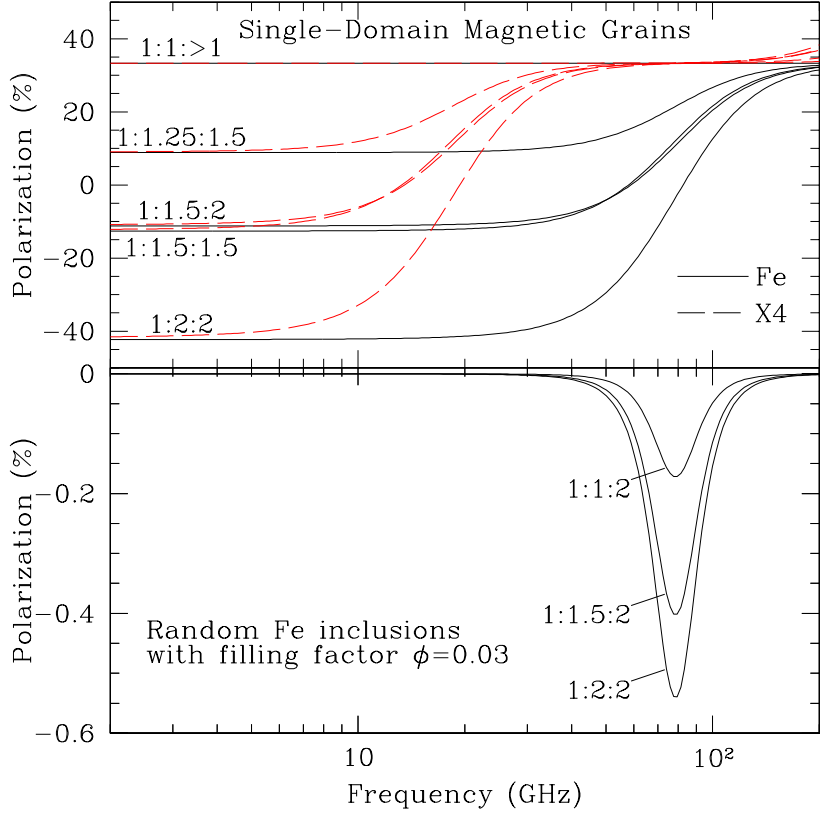}
\caption{\small  Polarization from magnetic grains. Upper panel:
Polarization of thermal emission from perfectly aligned single
domain grains of metallic Fe (solid lines) or hypothetical magnetic
material that can account for the Foreground X (broken lines).
Lower panel: Polarization from perfectly aligned grains with
Fe inclusions (filling factor is 0.03). Grains are ellipsoidal and
the result are shown for various axial ratios. From Draine \& Lazarian 1999.}
\label{micromag}
\end{figure}

The predictions for the polarization arising from magnetic grains  are shown in Fig.~\ref{micromag}

\subsection{Synergy of CMBPol and grain alignment studies}

Understanding of grain alignment is essential for the successful separating of the CMB polarization from
the polarization arising from dust. Any construction of a polarization template for the dust foreground uses the
assumptions on grain alignment.

The progress in understanding of grain alignment can be gauged by modeling of polarization in interstellar gas, molecular clouds,
accretion disks. The correspondence of the theory and observations obtained along these directions is encouraging (see Lazarian 2007).
However, more detailed modeling is required. We hope that by the time of CMBPol mission the grain alignment theory will be better tested
and therefore reliable. However, additional tests that CMBPol data can provide will also be useful. For the purposes of testing grain alignment
the larger coverage of frequencies is advantageous. Thus the project EPIC-CS or EPIC-2m would be preferable. Quantitative predictions of grain alignment 
that are possible these days (see \S\ref{predictions}) allow quantitative testing.  

In other words, CMBPol requires better understanding of grain alignment to get cleaner polarization signal from the early Universe and grain alignment
theory can benefit from the CMBPol mission. The importance of the latter should not be underestimated.
Indeed, grain alignment theory is aimed at studies of magnetic fields in various environments. In
Lazarian (2007), apart from interstellar medium, other environments, e.g. dark clouds, accretion disks, comets,
galactic nuclei, AGNs and Seyfet galaxies, are discussed. Therefore the progress in understanding how well magnetic fields are traced by aligned dust particles is essential for many astrophysical studies.

Combining polarimetric data at very different wavelength it is possible to
get insight into magnetic field structure. Examples of such studies combining far infrared, near infrared and
optical polarimetry are discussed in Lazarian (2007). In addition, one may mention yet another interesting
possibility. The studies of molecular cloud column densities with
the near infrared scattered light were presented in Padoan et al. (2006)
and Juvela et al. (2006). Those have shown that large scale mapping of scattered
intensity is possible up to $A_v\sim 10$ even for clouds illuminated by the average interstellar
radiation field. The polarization of scattered light should be affected by
grain alignment. This opens interesting prospects of detailed mapping of magnetic
fields at sub-arcsecond resolution. 

{\bf Acknowledgments}.
I thank Hoang Thiem for help in preparing the review. 
I acknowledge the support by the NSF grant  AST-0507164,
as well as by the NSF Center for Magnetic Self-Organization in Laboratory and Astrophysical
Plasmas.

\end{document}